\documentclass[iop,numberedappendix]{emulateapj}

\newcommand{\htdp}{H$_2$D$^+$}
\newcommand{\ntdp}{N$_2$D$^+$}
\newcommand{\nthp}{N$_2$H$^+$}
\newcommand{\hdtp}{HD$_2^+$}
\newcommand{\be}{\begin{equation}}
\newcommand{\ee}{\end{equation}}
\newcommand{\ba}{\begin{align}}
\newcommand{\ea}{\end{align}}

\usepackage{natbib}
\usepackage{enumerate}
\usepackage{color}
\usepackage{threeparttable}
\usepackage{tablefootnote}
\usepackage{amsmath}

\usepackage{subfigure}
\usepackage{graphicx}

\begin{document}

\title{Exclusion of Cosmic Rays in Protoplanetary Disks. II. Chemical Gradients and Observational Signatures}
\shorttitle{Exclusion of Cosmic Rays in Disks. II.}
\author{L. Ilsedore Cleeves\altaffilmark{1}, Edwin A. Bergin\altaffilmark{1} and Fred C. Adams\altaffilmark{1,2}}

\altaffiltext{1}{Department of Astronomy, University of Michigan, 1085 S. University Ave, Ann Arbor, MI 48109}
\altaffiltext{2}{Department of Physics, University of Michigan, 450 Church St, Ann Arbor, MI 48109}

\begin{abstract}
The chemical properties of protoplanetary disks are especially sensitive to their ionization environment.  Sources of molecular gas ionization include cosmic rays, stellar X-rays and short-lived radionuclides, each of which varies with location in the disk. This behavior leads to a significant amount of chemical structure, especially in molecular ion abundances, which is imprinted in their submillimeter rotational line emission. Using an observationally motivated disk model, we make predictions for the dependence of chemical abundances on the assumed properties of the ionizing field. We calculate the emergent line intensity for abundant molecular ions and simulate sensitive observations with the Atacama Large Millimeter/Sub-millimeter Array (ALMA) for a disk at $D=100$~pc. The models readily distinguish between high ionization rates ($\zeta\gtrsim10^{-17}$~s$^{-1}$ per H$_2$) and below, but it becomes difficult to distinguish between low ionization models when $\zeta\lesssim10^{-19}$~s$^{-1}$.  We find that \htdp\ emission is not detectable for sub-interstellar CR rates with ALMA (6h integration), and that \ntdp\ emission may be a more sensitive tracer of midplane ionization. HCO$^+$ traces X-rays and high CR rates ($\zeta_{\rm{CR}}\gtrsim10^{-17}$~s$^{-1}$), and provides a handle on the warm molecular ionization properties where CO is present in the gas. Furthermore, species like HCO$^+$, which emits from a wide radial region and samples a large gradient in temperature, can exhibit ring-like emission as a consequence of low-lying rotational level de-excitation near the star.  This finding highlights a scenario where rings are not necessarily structural or chemical in nature, but simply a result of the underlying line excitation properties.
 \end{abstract}

\keywords{astrochemistry --- circumstellar matter --- protoplanetary disks --- radiative transfer --- stars: pre-main sequence}  

\section{Introduction} 

Ionization plays an important role in setting thermal \citep[e.g.,][]{glassgold2004}, dynamical \citep{balbus1991,gammie1996,matsumura2003}, and chemical \citep[e.g.,][]{semenov2004,oberg2011a} properties of protoplanetary disks.  The chemistry occurring in the bulk mass of disks is especially sensitive to ionization for two reasons: (i) in the cold gas, ion-neutral reactions are the most efficient means towards chemical complexity, and (ii) in the ices, the crucial hydrogenation reactions \citep{tielens1982,hhl} are fueled by hydrogen atoms that are extracted by ionization of molecular H$_2$. The dominant ionization processes in disks are photoionization from stellar and interstellar UV and X-ray radiation, thermal ionization, ionization by the decay products of short-lived radionuclides (SLRs), and cosmic ray (CR) ionization.  Additionally, the cluster environment can provide a source of ionization on the outer surface of the disk from interstellar FUV, which can exceed that of the galactic average interstellar FUV by a factor of $\gtrsim3000$ \citep{fatuzzo2008}. 
Of these sources, CRs and potentially SLR decay provide ionization in the densest and coldest layers of the disk, where UV and X-rays are highly attenuated.  However, it is unclear as to whether or not CRs are actually present at rates derived for the interstellar medium  (ISM), i.e., $\zeta_{\rm{CR}}\sim(1-5)\times10^{-17}$~s$^{-1}$ in dense gas and $\zeta_{\rm{CR}}\sim(1-8)\times10^{-16}$~s$^{-1}$ in the diffuse ISM \citep[][and references therein]{dalgarno2006}.  Modulation of the CR flux can occur as a result of exclusion by natal stellar winds as explored in detail in \citet[][hereafter Paper I]{cleeves2013a} and discussed in \citet{glassgold1999,aikawa1999} and \citet{fromang2002}, in addition to exclusion by magnetic fields \citep{dolginov1994,padovani2011,fatuzzo2014}.  At these reduced levels, the ionization from SLR decay products becomes as important, and perhaps even more so, than that of CRs.  

In the present work, we set out to determine the chemical imprint of individual ionization processes in a generic protoplanetary disk model.  We outline how observations of molecular species can be used as a blueprint to constrain the underlying ionization properties of the disk's dense molecular gas.  More specifically, we focus on the impact of the assumed CR flux on molecular abundances in tandem with a detailed treatment of ionization by SLRs and stellar X-rays, including a Monte Carlo treatment of the scattered X-ray radiation field.  We extend these results to make testable predictions for future sensitive observations with the Atacama Large Millimeter/Submillimeter Array (ALMA) of molecular ion emission in protoplanetary disks.  Such predictions will help more accurately determine not only the ionization fraction in disks, which is important for constraining models of turbulence and chemistry, but also the {\em source} of ionization traced by a given molecular ion and transition.  

Current detections of molecular ions in disks include N$_2$H$^+$, HCO$^+$, and DCO$^+$ \citep[e.g.,][]{dutrey1997,vandishoeck2003,thi2004,oberg2010,oberg2011b}. The elusive \htdp\ has yet to be detected in a disk, with reported detections of the ($1_{10}-1_{11}$) line towards DM Tau and TW Hya \citep{ceccarelli2004} not confirmed by reanalysis or more sensitive observations \citep{guilloteau2006,qi2008,chapillon2011}. The \htdp\ upper limits nonetheless provide some constraints on the midplane ionization rate.  \citet{chapillon2011} investigated chemical models incorporating deuterium chemistry and find models with midplane ionization rates below $\zeta_{\rm total}<3\times10^{-17}$~s$^{-1}$ are required to explain the line's non-detection.  Strictly speaking, CRs are the primary midplane ionizing agent considered by \citet{chapillon2011}, though we note this value more generally provides a limit on the {\em total} ionization rate, which is expected to be primarily due to SLRs and CRs in the \htdp\ emitting gas, discussed in more detail below.  We note that the deuterium chemistry of \htdp, specifically the reactions leading back to H$_3^+$, are sensitive to the ortho-to-para ratio of both H$_3^+$ and H$_2$ \citep{chapillon2011,hugo2009,albertsson2014}, which we include in this work (see \S\ref{sec:chemmod}).  Furthermore, \htdp\ can undergo deuterium substitution towards the fully substituted and unobservable D$_3^+$ \citep{walmsley2004}, which complicates the interpretation of \htdp\ measurements in determining ionization from \htdp.   Such progressive D-substitution may point towards future difficulty in detection experiments of what would otherwise be a useful tracer of ionization and cold chemistry.  This point further emphasizes the utility of chemical models in the interpretation of molecular ion emission as a proxy for measuring disk ionization rates.

The paper is structured as follows:  In Sections \S\ref{sec:str} and \S\ref{sec:ion}, we review the physical model and the sources of ionization considered.  In Section~\S\ref{sec:chemmod} we describe the updated chemical reaction network used to predict the molecular abundances, and Section \S\ref{sec:chemres} examines the chemical abundance variations between the different ionization models.  Section \S\ref{sec:emit} makes predictions for sensitive, high-spatial resolution submillimeter observations that can be used to determine and to distinguish between the important ionization mechanisms. In Section \S\ref{sec:fc}, we discuss the effects of a more X-ray luminous source, time-decay of radionuclides, disk gas mass, and the impact of disk magnetic fields. In Section \S\ref{sec:discussion}, we summarize our results and discuss their implications.  
\begin{figure*}[ht!]
\centering
\includegraphics[width=0.99\textwidth]{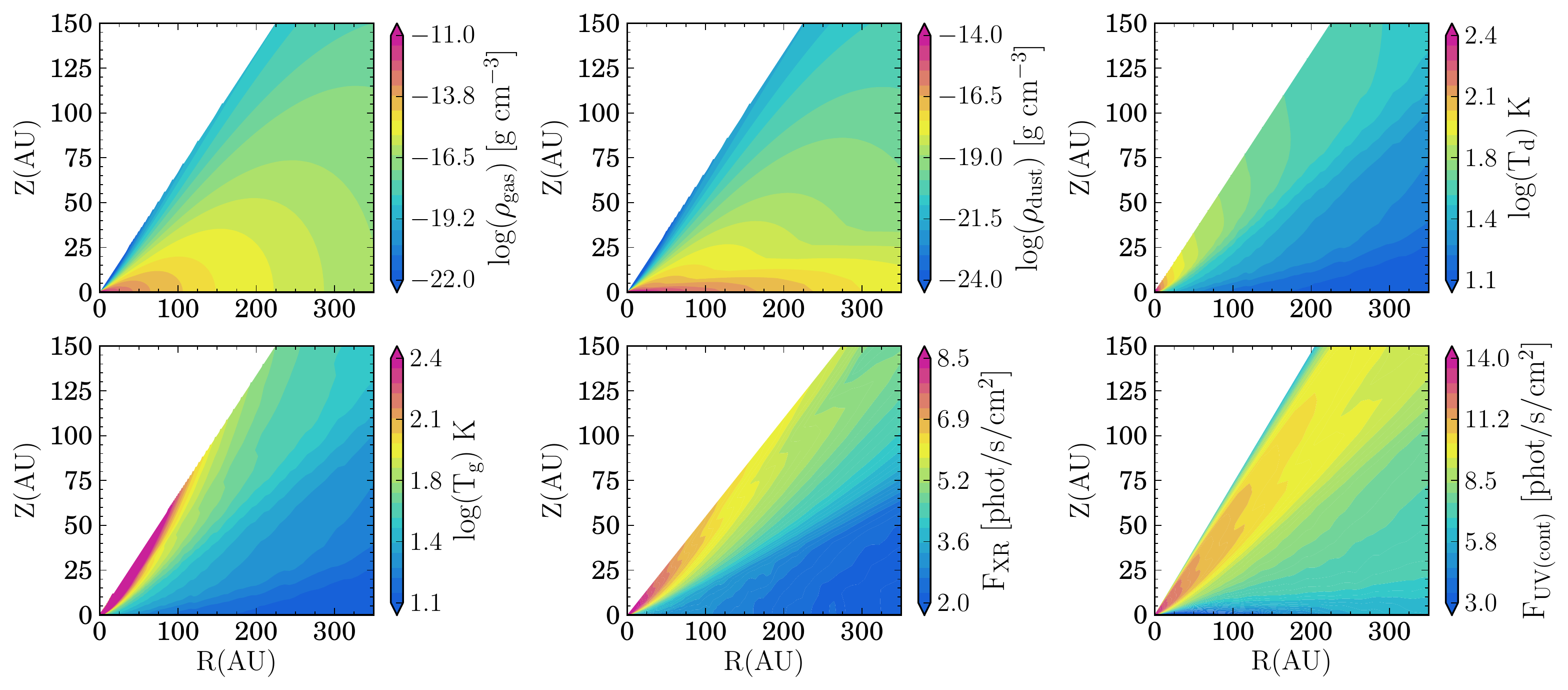}
\caption{Disk model used for the chemical calculations. {\em Top row, left to right:} i) gas density, ii) settled dust density, iii) dust temperature. {\em Bottom row, left to right:} i) gas temperature, ii) integrated X-ray radiation field and iii) FUV radiation field from the central star. The FUV flux is integrated between $930-2000$~\AA\ using the observed TW Hydra FUV spectrum \citep{herczeg2002,herczeg2004}, including Ly-$\alpha$.  The X-ray luminosity is $L_{\rm XR}=10^{29.5}$~erg~s$^{-1}$ integrated between $E_{\rm XR}=0.1-10$~keV.  \label{fig:str}}
\end{figure*}

\section{The Physical Model}
\label{sec:str}
We refer the reader to a detailed description of the disk model parameters in Paper I, and review only the main aspects of the model below.  The fiducial disk mass is $M_d\sim0.04$~M$_\odot$ within a $R_d=400$~AU radius, where the vertical structure and geometrical parameters (i.e. flaring) are typical of observed protoplanetary disks \citep{andrews2011}.  The gas and dust surface densities follow a power law with an exponential taper at the outer edge \citep{hughes2008,andrews2011} and the dust is heated via passive irradiation from the central K5V star with T$_{\rm eff}$ = 4300 K with the code TORUS \citep{harries2000,harries2004,kurosawa2004,pinte2009}.  The gas temperature is calculated from a fitting function calibrated to detailed thermochemical models of disks heated by FUV photons from the central star (Simon Bruderer in private communication, 2013).  In the heating calculation, we consider only the central star's FUV field (described in \S\ref{sec:photo}). The vertical distribution of dust is broken up into two populations to simulate grain growth, with a population of large millimeter grains concentrated at the geometrical center of the disk (the midplane) and a population of ``atmosphere'' micron-sized grains distributed over larger scale heights.  The gas and small micron-sized grains follow the same scale height (see Paper I for details, as well as the dust optical properties).  Figure~\ref{fig:str} shows the disk density, thermal and stellar radiation fields, and Table~\ref{tab:modpar} outlines the main physical parameters of our model.

\begin{deluxetable}{lr}
\tablecolumns{2} 
\tablewidth{0pt}
\tablecaption{Stellar and disk model parameters.\label{tab:modpar}}
\tabletypesize{\footnotesize}
\tablehead{ &                        
}
\startdata
  {\bf Stellar Model} \\ 
Stellar Mass & 1.06 M$_{\odot}$  \\
Stellar Radius & 1.83 R$_{\odot}$ \\ 
Stellar T$_{\rm eff}$ & 4300 K \\
L$_{\rm UV}$ &  2.9 $\times 10^{31}$ erg/s \tablenotemark{a} \\
L$_{\rm XR}$  & 10$^{29.5}$ erg/s\\
 \\
{\bf Disk Model} \\ 
 R$_{\rm inner}$  &  0.1 AU \\
  R$_{\rm outer}$  &  400 AU \\
M$_{\rm dust}$  &    3.9 $\times 10^{-4}$ M$_{\odot}$ \\ 
M$_{\rm gas}$  &    0.039 M$_{\odot}$ 
\enddata
\tablenotetext{a}{As computed from the observed FUV spectrum of TW Hya integrated between 930 and 2000 \AA\ \citep{herczeg2002,herczeg2004}.}
\end{deluxetable}

\section{Ionization Sources}
\label{sec:ion}

Star/disk systems are subject to a variety of ionization sources, including FUV and X-ray radiation from the central stars, short-lived
radionuclides, and CRs.  These ionization sources are not constant in time or in space. Instead, they are expected to vary with the local environment, from system to system, and display time dependence.  Some environments include ionizing
radiation from nearby (more massive) stars \citep[see the review of][]{adams2010}, although this complication
is left for future work. This section outlines the physical mechanisms that
contribute to the disk ionization rate and the ranges of values considered at present.

\subsection{Stellar Photoionization}\label{sec:photo}
T Tauri stars are known to be X-ray luminous with X-ray luminosities typically ranging $L_{\rm{XR}}\sim10^{29}-10^{31}$~erg~s$^{-1}$ \citep[e.g.,][]{feigelson1981,feigelson1993,telleschi2007}. Consequently, the disk's X-ray ionization properties are perhaps the best (observationally) constrained ionization agent present in the disk molecular gas.  That said, the permeability of the disk gas to X-rays is unknown owing to uncertainties in the disk mass (the column density) and composition (opacity) of the absorbing materials.  The biggest uncertainty in the X-ray opacity is in the time-dependent effects of dust settling, which redistributes and removes the absorbing material from the upper layers.  More specifically, by removing dust mass from the upper layers the opacity shifts from a well-mixed gas and dust regime to a gas-dominated opacity.  For \citet{asplund2009} abundances, the change in opacity corresponds to a factor of two decrease in absorption cross section between the well-mixed and fully settled cases at $E_{\rm XR}=1$~keV \citep{bethell2011a}.    However, with knowledge of the star's X-ray luminosity, it may be possible to shed light on the permeability of the disk atmosphere to X-rays with the proposed molecular tracers in this work (see \S\ref{sec:chemres}). 

We calculate the stellar FUV and X-ray radiation fields as a function of position and wavelength within the disk using a Monte Carlo treatment described in \citet{bethell2011b}.  For FUV transfer, we consider the dust model's position-dependent opacities and compute the absorption and scattering on dust grains.  In addition to the radiative transfer through the dust, we calculate the Lyman-$\alpha$ line transfer, where photons scatter isotropically off atomic hydrogen present at the disk surface.  Treatment of Lyman-$\alpha$ is important as this line has been observed to carry $\sim 80-90\%$ of the total FUV flux \citep{herczeg2004,schindhelm2012,france2014}; as a consequence of its scattering properties, the Lyman-$\alpha$ photons are able to penetrate deeper into the disk gas than the rest of the primarily dust-scattered FUV photons \citep{bethell2011b}.  Because the present paper is mainly focused on understanding the ionization properties of the dense molecular gas, we hold the FUV radiation field constant in the models presented here since FUV mainly contributes to the ionization in the lower density surface layers where, for example, C$^+$ emission originates.  The assumed incident FUV spectrum is that of TW Hydra \citep{herczeg2002,herczeg2004}, the closest and least extincted T Tauri star along the line of sight, $d=55$~pc \citep{perryman1997}.  In this work, we do not include the interstellar FUV radiation field as a source of ionization in the dense gas, but it is expected to play a role in the cluster environment where the external FUV field can be many thousands of $G_0$ \citep{fatuzzo2008}, where $G_0$ is the mean value for the ISM \citep{habing1968,draine1978}. A $G_0$=1 has a less significant effect on the chemistry since a small amount of UV opacity from small grains restricts the UV penetration to an outer shell on the disk surface, which is dwarfed by scattered UV from star itself \citep[see Fig.~3 of][]{cleeves2013a}.  A similar ionization study should be carried out for more extreme cluster environments, where the high UV field will be accompanied by a higher SLR abundance and potentially CR abundance in the vicinity of massive stars.  

The specific (energy-dependent) X-ray fluxes are calculated using the same code as was used for the FUV \citep{bethell2011b} where we instead adopt the updated X-ray absorption gas and dust opacities of \citet{bethell2011a}.  We note that the model has been updated since Paper I, where the dust-reduced opacity in the upper layers (90\% reduction in dust, or ``e0p1'') was originally assumed to be uniform throughout the disk.  We have since updated this calculation to take into account our knowledge of the local gas-to-dust mass ratio. We determine the absorption opacity due to gas and dust individually at each location in the disk from the \citet{bethell2011a} cross sections.  The X-ray scattering is dominated by Thomson scattering and approximately isotropic (i.e., photons are scattered equally in the backwards and forwards directions).  For the primary model, we assume a characteristic T Tauri star X-ray luminosity of $L_{\rm{XR}}$ = 10$^{29.5}$~erg~s$^{-1}$ with the quiescent X-ray spectral template given in Paper I.  

\begin{figure}[h!]
\begin{centering}
\includegraphics[width=0.48\textwidth]{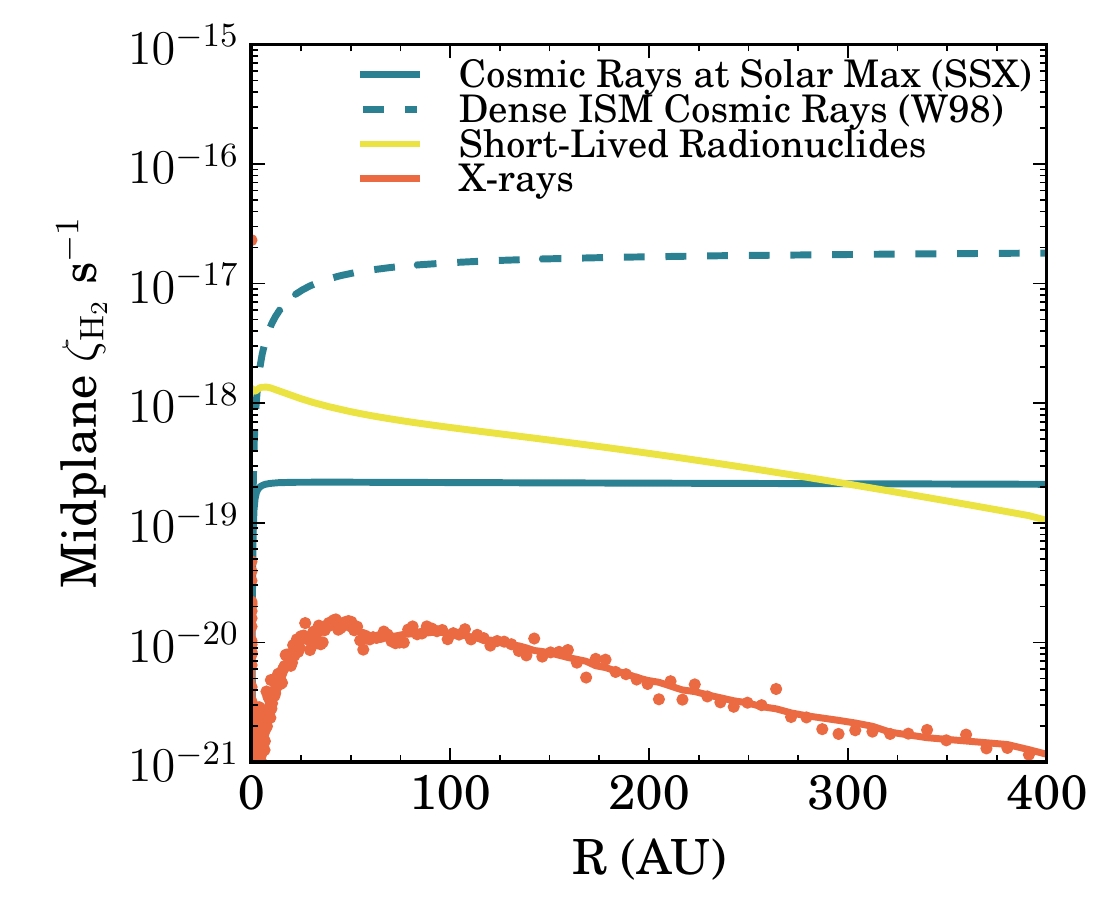}
\caption{Sources of H$_2$ ionization present in the midplane as a function of disk radius.  The orange points are the result of our Monte Carlo calculation and are dominated by intermediate energy X-ray photons, typically $5-7$~keV.  The solid blue-green line is the cosmic ray ionization rate including the modulation of stellar winds for the present-day Sun at solar maximum.  The CR value typically assumed for disk chemical models is shown in dashed blue-green for comparison.  The yellow line shows the initial contribution from short-lived radionuclide ionization, predominantly $^{26}$Al.  The radial decrease in SLR ionization is due to losses of SLR decay products, which escape the disk.  The effective half-life of the SLR curve is $t_{\rm{half}}\sim1$~Myr.     \label{fig:ions}}
\end{centering}
\end{figure}

The most abundant low energy X-rays ($E\sim1$~keV) are also the most easily attenuated, and consequently do not contribute to the dense midplane ionization.  The most important X-rays in the densest gas are typically of intermediate energies, between $5-7$~keV, which are less numerous but can more readily penetrate the gas and dust.  We emphasize that a detailed treatment of X-ray scattering is crucial in understanding disk ionization, otherwise there would be no X-ray photons in the midplane \citep[][Paper I]{igea1999,ercolano2013}.  Correspondingly, the scattered X-ray radiation field provides the absolute floor to the midplane ionization rate in the absence of CRs and SLRs (see Figure~\ref{fig:ions}).  In the Figure, the initial rise in midplane X-ray intensity occurs as the disk surface density drops (becomes more optically thin), while the fall-off beyond $R\sim50$~AU is simply geometrical dilution.  The  X-ray ionization floor falls in the range $\zeta_{\rm{XR}}\sim(1-10)\times10^{-21}$~s$^{-1}$.  For a more massive or denser disk than the one considered here, the role of X-rays can be diminished in the midplane by more extreme gas extinction.  On the other hand, a star with higher X-ray luminosity would have a proportionally higher $\zeta_{\rm{XR}}$ depending upon the shape of the stellar X-ray spectrum (see also \S\ref{sec:xray}).

\subsection{Short-Lived Radionuclide Ionization}\label{sec:slr}
The disk ionization contribution from the decay of SLRs has been studied extensively \citep{consolmagno1978,umebayashi1981,finocchi1997,umebayashi2009,umebayashi2013,cleeves2013b}.  The initial abundances of radioactive species within a typical protoplanetary disk are uncertain but can be estimated for the case of the protosolar disk from isotopic abundance measurements in meteorites \citep[e.g.,][and references therein]{wasserburg2006}.  Whether these values are representative of all disks is unknown; however,  the frequency of differentiated extrasolar asteroids seem to indicate that the young Solar System was at least not atypical in its SLR content \citep{jura2013}, a hypothesis which models successfully reproduce \citep{vasileiadis2013}.  Moreover, recent work indicates that both direct (disk) SLR injection in clusters and distributed SLR enrichment in molecular clouds can produce abundances comparable to those inferred for the early Solar Nebula \citep{adams2014}.

Time evolution of abundances adds further uncertainty in estimating the SLR contribution to the ionization budget in disks.  The characteristic half-life of the ensemble of important SLRs, primarily $^{26}$Al \citep[$t_{\rm{half}}=0.74$~Myr;][]{schramm1971,macpherson1995,umebayashi2009} and $^{60}$Fe \citep[$t_{\rm{half}}=2.6$~Myr\footnote{We note that in \citet{cleeves2013b}, the value assumed for the $^{60}$Fe half-life predates \citet{rugel2009}, where the original value of $t_{\rm{half}}=1.5$~Myr used in Paper I originates from \citet{kutschera1984}.  Once the $^{26}$Al has been depleted after approximately $t>5$~Myr, $^{60}$Fe is the primary contributor to SLR ionization.  However at this epoch, the ionization attributed to scattered stellar X-rays begins to exceed that of the SLR contribution (see Fig~\ref{fig:ions}), and correspondingly the change in chemical properties or ionization fractions are not expected to be large for the updated half-life.  Nonetheless, we have included an updated Figure~4 from \citet{cleeves2013b} and new fits to the midplane ionization rate in Appendix~A of this manuscript.};][]{rugel2009} corresponds to approximately $t_{\rm{half}}\sim1.2$~Myr (Appendix~A).  This implies that for disks aged 5~Myr, the contribution from SLRs is reduced by nearly 80\%, and consequently scattered X-rays and SLRs would contribute to the midplane ionization at a similar magnitude  outside of $R>50$~AU (Fig.~\ref{fig:ions}). Inside this radius, the SLR contribution exceeds that of X-rays due to the high attenuation of X-ray photons in the inner disk midplane.

Another complication is that a substantial fraction of the SLR decay products, e.g., beta particles and $\gamma$-rays, escape the disk prior to ionizing the gas when surface densities drop below $\Sigma_g\sim1-10$~g~cm$^{-2}$ so that not all of the available energy is deposited locally.  The present model incorporates the effects of $\gamma$-ray, $\beta^{+}$ and $\beta^{-}$ loss in the outer disk \citep[described in][]{cleeves2013b} for the settled disk slab model.  We note that the main chemical model results presented in \S\ref{sec:chemres} assume the SLR ionization rate is constant with time, which is acceptable for disks with $t<1$~Myr or less.  We relax this requirement in \S\ref{sec:rntime} where we consider SLR time evolution within the chemical calculation itself.

\subsection{Cosmic Ray Ionization}\label{sec:cr}
Ionization by cosmic rays at the interstellar rate ($1-5\times10^{-17}$~s$^{-1}$) is a commonly assumed ingredient in models of disk chemistry and physics.  However, detection limits of H$_2$D$^+$ emission suggest that the ionization rate is actually {\em lower} than expected for the dense ISM, pointing to some variety of additional attenuation.  One possible explanation is CR exclusion by natal stellar winds, i.e, an analogue to the modern day ``Heliosphere.'' Within the Heliosphere, the solar wind strongly modulates CRs, especially those below $E_{\rm CR}\sim100$~MeV.  Pre-main-sequence stars such as T Tauri stars are significantly more magnetically active, have high rotation rates, and high mass loss rates, all of which may drive high levels of CR exclusion at all particle energies.  Models of the Gyr-old Sun show significant modulation even at late times \citep{svensmark2006,cohen2012}, corresponding to incident CR ionization rates in the range $\zeta_{\rm{CR}}\sim(3-100)\times10^{-22}$~s$^{-1}$. In Paper I, we present a simple model of a scaled-up Heliosphere, i.e. a ``T-Tauriosphere,'' to estimate the degree to which a T Tauri star could potentially exclude CRs with winds, and how this exclusion is imprinted on the ionization state of the disk.  In the instance of relatively mild modern-day solar winds at solar minimum, the ionization rate by CRs at the disk surface is reduced to just $\zeta_{\rm{CR}}\lesssim10^{-18}$~s$^{-1}$.   This value is already over an order of magnitude below the typically assumed (dense gas) interstellar rate.   At solar maximum, the ionization rate at the disk surface is reduced to $\zeta_{\rm{CR}}\sim10^{-19}$~s$^{-1}$ (see Figure~\ref{fig:ions}). If young, energetic stellar winds are {\em even more efficient} at sweeping away CRs, the CR ionization rate will drop below the solar maximum value at which point SLRs now become the primary midplane ionization contributor, where $\zeta_{\rm{SLR}}\sim(1-10)\times10^{-19}$~s$^{-1}$ for $t<1$~Myr.  

While the ``T-Tauriosphere'' models explored in Paper I were simple scaled up versions of the modern day solar wind, the spectra reflect the general behavior we expect under more extreme stellar wind conditions such as those explored in \citet{cohen2012}, where low energy CRs are excluded most severely, and high energy CRs ($E>1$~GeV) are only weakly modulated.  Here we explore two wind-modulated CR ionization models, including the present-day Sun (Solar System Maximum: SSX) and a scaled-up exclusion model (T Tauri Maximum: TTX) as presented in Paper I. These models provide a realistic framework that allows us to quantify how the chemistry is affected by modulated incident CR fluxes.  We note that the wind modulation affects the incident spectrum and consequently the ionization rate {\em at the disk surface}, in addition to attenuation by the disk gas, where the typical attenuating surface density is $\Sigma_g\sim100$~g~cm$^{-2}$ \citep{umebayashi1981}.  Additional opacity can arise from magnetic mirroring and magnetic irregularities in the disk, which we discuss further in \S\ref{sec:discussion} \citep[see also][]{dolginov1994,padovani2011,fatuzzo2014}.

For the scaled up (TTX) CR model discussed in Paper I, the ionization rate due to CR drops below that of X-rays {\em at all radii}, and the TTX models correspond to a purely X-ray dominated disk for models without SLR ionization.  Table~\ref{tab:crrates} provides characteristic rates in the disk surface layers for the different CR wind modulation models considered in this work.  In addition to the wind modulated SSX and TTX spectra, we examine two ``ISM-like'' models, M02 \citep{moskalenko2002} and W98 \citep{webber1998}.  W98 is the closest model to what is typically assumed in models of disk chemistry, $\zeta_{\rm{CR}}\sim2\times10^{-17}$~s$^{-1}$, while the M02 ionization rate is characteristic of CR ionization rates measured from H$_3^+$ in the diffuse ISM \citep{indriolo2012}, which are unattenuated.  For disks without a significant amount of surrounding nebulous gas, such as TW Hya, the M02 model may be more representative of the interstellar (completely unmodulated) CR rate.  

\begin{deluxetable}{lcrrr}
\tablecolumns{2} 
\tablewidth{0pt}
\tablecaption{CR model ionization rates for $N({\rm H_2}) \le 10^{25}$ cm$^{-2}$. \label{tab:crrates}}
\tabletypesize{\footnotesize}
\tablehead{ Model &  ID &  $\zeta_{\rm CR}$  }
\startdata
 \citet{moskalenko2002}  & M02 & $ 6.8\times 10^{-16}$ s$^{-1}$\\
\citet{webber1998} & W98 & $ 2\times 10^{-17}$ s$^{-1}$\\
Solar System Maximum  &SSX & $ 1.6 \times 10^{-19}$ s$^{-1}$\\
T Tauri Maximum & TTX & $ 1.1 \times 10^{-22}$ s$^{-1}$
\enddata
\end{deluxetable}

\section{The Chemical Network} 
The chemical nature of a parcel of interstellar gas and ice is highly sensitive to the properties of its environment.  These properties include the radiation field intensity, grain size and volume density (regulating freeze-out and desorption), gas/dust temperatures and densities.  In cold gas ($T<100$~K), the gas-phase chemistry is particularly sensitive to the gas ionization fraction, as ion-neutral reactions are the most important (i.e., quickest) gas phase reactions that take place \citep{herbst1973}. The geometry of disks, ranging between lower density, warm ionized surfaces to predominantly neutral, cold midplanes provides a diversity of physical conditions, which are directly translated into a rich chemical environment.  Conversely, observations of the chemical composition of disks provide clues into the underlying physical environment and therefore are a powerful observational tool to help understand the important physics governing these systems.  In the following section we examine how the chemical properties of disks, particularly the molecular ions, react to different assumptions regarding ionization processes, and how they may be used as diagnostic tools.

\subsection{Chemical Model}\label{sec:chemmod}
The physical model and stellar UV and X-ray radiation fields provide the backbone on which we solve for the time-dependent chemical abundances with the \citet{fogel2011} disk chemistry code.  This chemical reaction network is based upon the OSU gas-phase network \citep{smith2004}, where \citet{fogel2011} substantially expanded the network to include important processes such as thermal and non-thermal sublimation, photodissociation, freeze-out onto grains, CO and H$_2$ (and isotopologues) self-shielding, and stellar and non-stellar ionization of H$_2$ and helium.  The code is calculated as 1+1D, where different disk radii are treated independently and self-shielding is considered in the vertical direction.  The code is run in parallel with the publicly available GNU Parallel software \citep{Tange2011a}. 
 We note that in the calculation of the temperature structure and UV transfer, we consider the spatial dependence of the grain size populations in detail.  The grain-surface chemistry, however, is fixed to a single ``typical'' grain-size of $r_g=0.1\mu$m, where the underlying assumption is that the small grains dominate the surface area and are most important for the chemistry.  Because the gas and small grains are uniformly distributed, this approximation is justified in the present work; however, we would nonetheless be ``missing'' surface area in the midplane where the 1mm grains are concentrated.  We note, though, that this correction should be small considering that small grains are present throughout the disk at all scale-heights.  We furthermore note that the physical size of $r_g$ is not the important quantity, but rather this number translates into an effective grain surface area per unit volume in the chemical code (e.g., $7.5\times10^{-3}$~$\rm \mu m^2$~cm$^{-3}$ at $n_{\rm H_2}=10^{10}$~cm$^{-3}$).  Nonetheless, future work should explore more than two size populations, including their vertical distribution in the disk \citep[for example, in the formalism of][]{dullemond2004} and how these will affect the grain-chemistry.  

In order to make predictions for deuterated molecular ions, we have added to the \citet{fogel2011} network a simple deuterium chemistry to predict the abundances of H$_2$D$^+$ and N$_2$D$^+$, which become enriched relative to the main isotopologue due to a chemical favorability towards the heavier isotopologue at low ($T<50$~K) temperatures \citep{millar1989}.  Even though DCO$^+$ is part of the network through the \htdp+ CO formation pathway, we do not make predictions for the DCO$^+$ abundances in the present work; the chemistry of DCO$^+$ depends sensitively on the deuterated carbon chemistry, for which we have not included a complete network.  Instead, the network is designed to reliably predict the relatively simpler H$_2$D$^+$ and closely related N$_2$D$^+$ chemistry.   An important facet of the \htdp\ chemistry is that the reaction rates depend strongly on the ortho-to-para ratio of the reactants, H$_2$ and H$_3^+$, and their isotopologues \citep{pagani2009,hugo2009,chapillon2011}.  We approximate this behavior by assuming that the ortho-to-para ratio is locally {\em thermal} for H$_2$D$^+$ and include the ortho-to-para fraction as weights on the net reaction rate.  In the following prescription we designate $f^1_o$ ($f^1_p$) as the fraction of H$_2$ in the ortho (para) state, and $f^2_o$ ($f^2_p$) as the fraction of \htdp\ in the ortho (para) state.  The weighted reaction rate coefficients for (o)rtho and (p)ara H$_2$ and \htdp\ is: 
\begin{align}
R(T) = &R^{\rm oo}(T)f^1_{\rm o}(T)f^2_{\rm o}(T)  + \nonumber  \\ 
&R^{\rm op}(T)f^1_{\rm o}(T)f^2_{\rm p}(T) + ... ,
\end{align}
where $ f^{1}_o + f^{1}_p = 1$, $ f^{2}_o + f^{2}_p = 1$, and $R^{\rm oo}(T)$ is the reaction rate coefficient for reactions between o-H$_2$ and o-\htdp, for example.   The o/p ratio of H$_2$ is given by
\ba\label{first}
{\rm o/p}_{\rm H_2}(T_{\rm{gas}})=&5.3534\times10^{-3}+ \nonumber  \\ 
&\frac{(3.0346-5.3534\times10^{-3})}{\left[1+(96.5330/T_{\rm{gas}})^{3.5096}\right]},
\end{align}
for gas at temperatures above $T_g\ge80$~K and for temperatures below that value,
\be\label{third}
{\rm o/p}_{\rm H_2}(T_{\rm{gas}})=9 \exp{(-170.5/T_{\rm{gas}})}.
\ee
\citet{flower2006} finds that at very low temperatures ($T_g\lesssim10$~K) the ortho-to-para ratio of H$_2$ exceeds the Boltzmann value, and to approximate this behavior we set a floor to the o/p ratio of H$_2$ at 10$^{-3}$ that limits Eq.~(\ref{third}) from dropping below this value.  The o/p ratio of H$_2$D$^{+}$ is given by
\ba\label{second}
{\rm o/p}_{\rm H_2D^+}(T_{\rm{gas}})=&-1.6977\times10^{-2}+ \nonumber  \\
&\frac{(3.0375+1.6977\times10^{-2})}{\left[1+(47.9640/T_{\rm{gas}})^{3.0692}\right]},
\end{align}
where both o/p formalisms in Eqs.~(\ref{first}) and (\ref{second}) are from \citet{lee2014}, submitted.  

In the chemical network, the deuterium extension to the reaction set -- for the most part -- mirrors the main isotopologues, and we assume statistical branching ratios where necessary.  While the ionization of H$_2$ and helium are the first step towards the molecular ion chemistry, we do allow for HD and D$_2$ to be directly ionized by X-rays, CRs and SLRs, assuming the same cross sections as for H$_2$. The self-shielding functions for HD and D$_2$ are from \citet{wolcottgreen2011}.  For the reactions where rates are known for the deuterated isotopologue, we incorporate values from the literature for H$_2$D$^+$ electron recombination and reactions with atomic H \citep{roberts2004}, H$^+$ and D$^+$ reactions with atomic/molecular hydrogen/deuterium, DCO$^+$, N$_2$D$^+$ reactions with H \citep{roberts2000}, and neutral-neutral warm water-formation deuterium reactions including their significant barriers \citep{bergin1999}.  While we include the spin information for the H$_2$ + H$_2$D$^+$ reaction rates, the endothermicity of the back reactions for the overall less abundant heavier isotopologues, HD$_2^+$ and D$_3^+$ are taken to be a single value in the network, 187~K and 341~K, respectively \citep{roberts2004}. In addition to the standard set of gas-phase reactions in the original network, we include a simple grain-surface chemistry allowing for H$_2$/HD/D$_2$, H$_2$O/HDO, and H$_2$CO/HDCO formation on grain surfaces through hydrogenation \citep{hhl} and CO$_2$ formation through reactions of CO with OH and O with barriers \citep{garrod2011}.   In total, the full network spans over $\sim6200$ reactions and $\sim600$ species.  

The initial chemical abundances are listed in Table~\ref{tab:abun} and are for the most part adopted from \citet{fogel2011} based upon the dark cloud model of \citet{aikawa1999a}.  The abundances of CS and SO have been adjusted to the observationally derived abundances of the TMC-1 dark cloud from \citet{bachiller1990} originally compiled in \citet{guelin1988} and \citet{rydbeck1980}.  
Together, CS and SO constitute the main sulfur reservoir with an abundance relative to hydrogen number totaling $\sim10^{-8}$.  The atomic (ionized) sulfur, S$^+$, is reduced to a low value, where we assume $10^{-11}$.  The motivation behind the total reduced sulfur abundance,  $\sim10^{-8}$ rather than the diffuse ISM value of $\sim10^{-6}$, comes from the observation that the volatile sulfur abundance in dense clouds is far lower than that of the diffuse ISM \citep{joseph1986,tieftrunk1994}.  These findings were confirmed by observations of molecular sulfur-bearing species including CS and SO \citep[e.g.,][]{langer1996,wakelam2004} where the molecular abundances measured in dense clouds are far less than those predicted by chemical calculations for a diffuse ISM sulfur abundance \citep{hatchell1998,wakelam2004}.  In particular, \citet{wakelam2011} find that based upon chemical modeling of dense (high mass) cores, the volatile sulfur abundance is found to be even further substantially reduced, where the observations account for an abundance of sulfur totaling $2\times10^{-9}-5\times10^{-8}$ relative to hydrogen across four sources.  The interpretation is that the sulfur has been converted into a more refractory form, i.e., a ``sulfur-rich residuum'' \citep{wakelam2004}, and is not chemically available to produce volatile sulfur-bearing molecules.  Evidence of high levels of atomic sulfur in shocked gas near Class 0 protostars further supports this scenario, where in one explanation the observed sulfur atoms are sputtered from the sulfur-rich residuum \citep{anderson2013}. Likewise, the ionized metal abundances of Si$^+$, Mg$^+$, and Fe$^+$ have been reduced to low levels ($10^{-11}$) based upon the chemical modeling results of \citet{maret2007} for the B68 pre-stellar core.   The HD abundance is based upon the protosolar bulk value, where D/H in HD is $2.0\pm0.35\times10^{-5}$ \citep{geiss2003}.  The \htdp, \hdtp, and D$_3^+$ abundances are estimated from \citet{walmsley2004} for the small (0.025~$\mu$m) grain case at high ($n_{\rm H_2}=10^7$~cm$^{-3}$) density.  Regarding the nitrogen abundances, we note that in these models, the initial abundance of nitrogen is primary atomic.  Had we begun with a larger fraction molecular nitrogen, the overall N$_2$H$^+$ abundance would increase but the shape of the column density profile remains largely unchanged (Schwarz \& Bergin 2014, submitted; Cleeves et al. 2014, in prep.).  Starting off with a larger fraction of ammonia, for example, tends to reduce the overall N$_2$H$^+$ abundance across the bulk disk (outside of the ammonia snowline) by removing nitrogen from the gas-phase.  Consequently, the shape of the column density profile of \nthp\ is still sensitive to the CR ionization rate even without additional information on the disk nitrogen abundances.  

\begin{deluxetable}{llll} 
\tablecolumns{4} 
\tablewidth{0pt}
\tablecaption{Initial chemical abundances, $\chi$. \label{tab:abun}}
\tabletypesize{\footnotesize}
\tablehead{ Species &  
log($\chi$) &
Species &
log($\chi$)                      
}
\startdata
H$_2$ & $5.00\times10^{-1}$ & H$_2$O(gr) & $2.50\times10^{-4}$ \\
He & $1.40\times10^{-1}$ & N & $2.25\times10^{-5}$ \\
CN & $6.00\times10^{-8}$ & H$_3^+$ & $1.00\times10^{-8}$ \\
CS & $4.00\times10^{-9}$ & SO & $5.00\times10^{-9}$ \\
Si$^+$ & $1.00\times10^{-11}$ & S$^+$ & $1.00\times10^{-11}$ \\
Mg$^+$ & $1.00\times10^{-11}$ & Fe$^+$ & $1.00\times10^{-11}$ \\
C$^+$ & $1.00\times10^{-9}$ & CO & $1.00\times10^{-4}$ \\
N$_2$ & $1.00\times10^{-6}$ & C & $7.00\times10^{-7}$ \\
NH$_3$ & $8.00\times10^{-8}$ & HCN & $2.00\times10^{-8}$ \\
HCO$^+$ & $9.00\times10^{-9}$ & H$_2$CO & $8.00\times10^{-9}$ \\
HD & $2.00\times10^{-5}$ & H$_2$D$^+$ & $1.30\times10^{-10}$ \\
 HD$_2$$^+$ & $1.00\times10^{-10}$ & D$_3$$^+$ & $2.00\times10^{-10}$ \\
  C$_2$H & $8.00\times10^{-9}$ &
\enddata
\end{deluxetable}

\subsection{Chemical Abundance Results}\label{sec:chemres}
\begin{figure*}[ht!]
\centering
\includegraphics[width=1.0\textwidth]{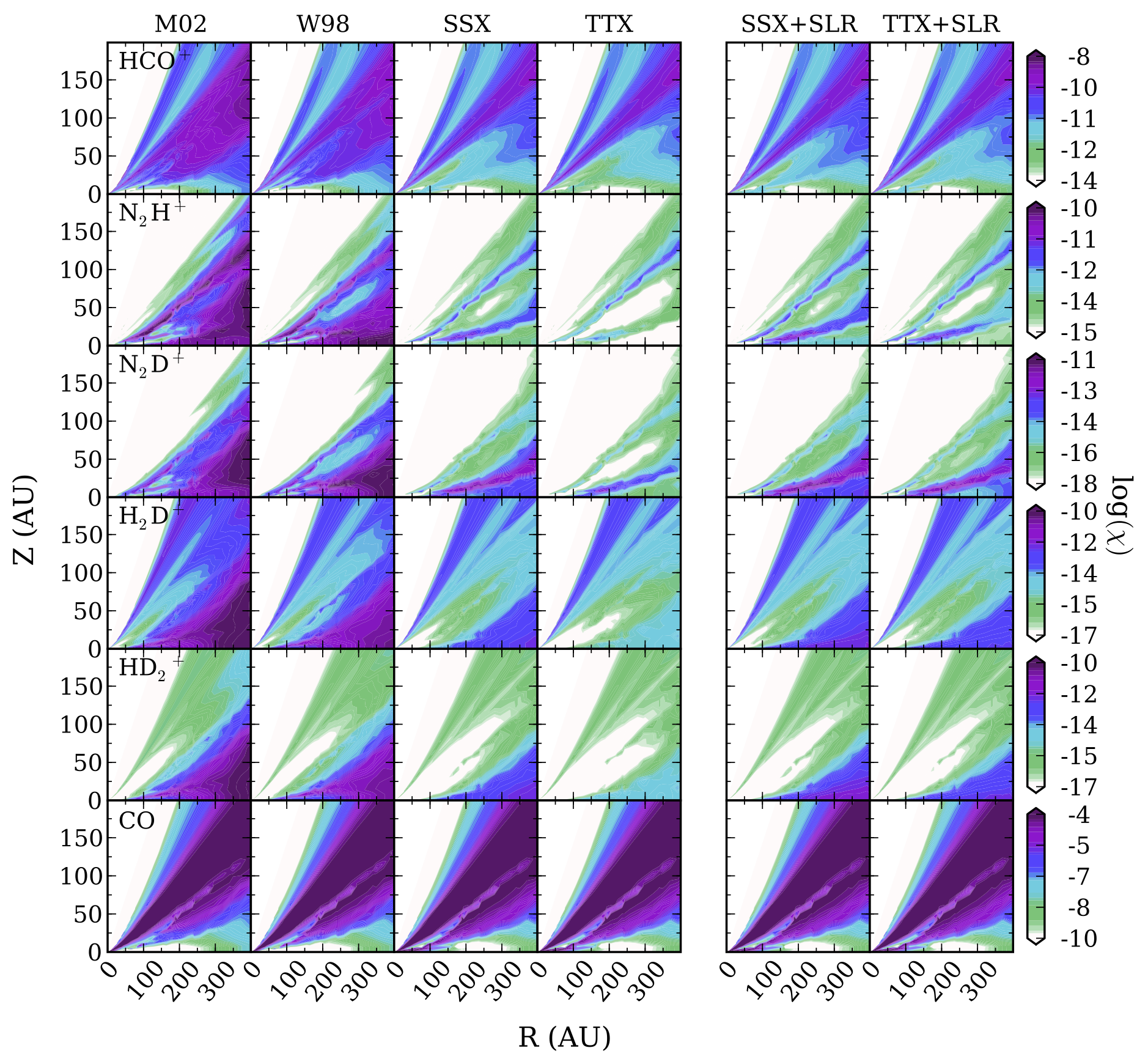}
\caption{Disk chemical abundances relative to H$_2$ as a function of radius and height, where $R,Z$ = (0,0)~AU is the location of the central star. Abundances are shown at time $t=1$~Myr of chemical evolution. Columns correspond to different CR/SLR ionization models for a fixed X-ray luminosity, and rows are different molecular abundances as labeled on the leftmost column.  Specific CR and/or SLR ionization model is indicated at the top of each column.  The two rightmost columns include contribution from the decay of short-lived radionuclides adding at most $\zeta_{\rm{SLR}}\sim10^{18}$~s$^{-1}$ to the total H$_2$ ionization (see Figure~\ref{fig:ions}).  Their inclusion provides a floor for the molecular ion abundance (see especially N$_2$H$^+$). \label{fig:abundance}}
\end{figure*}
\begin{figure}[ht!]
\centering
\includegraphics[width=0.45\textwidth]{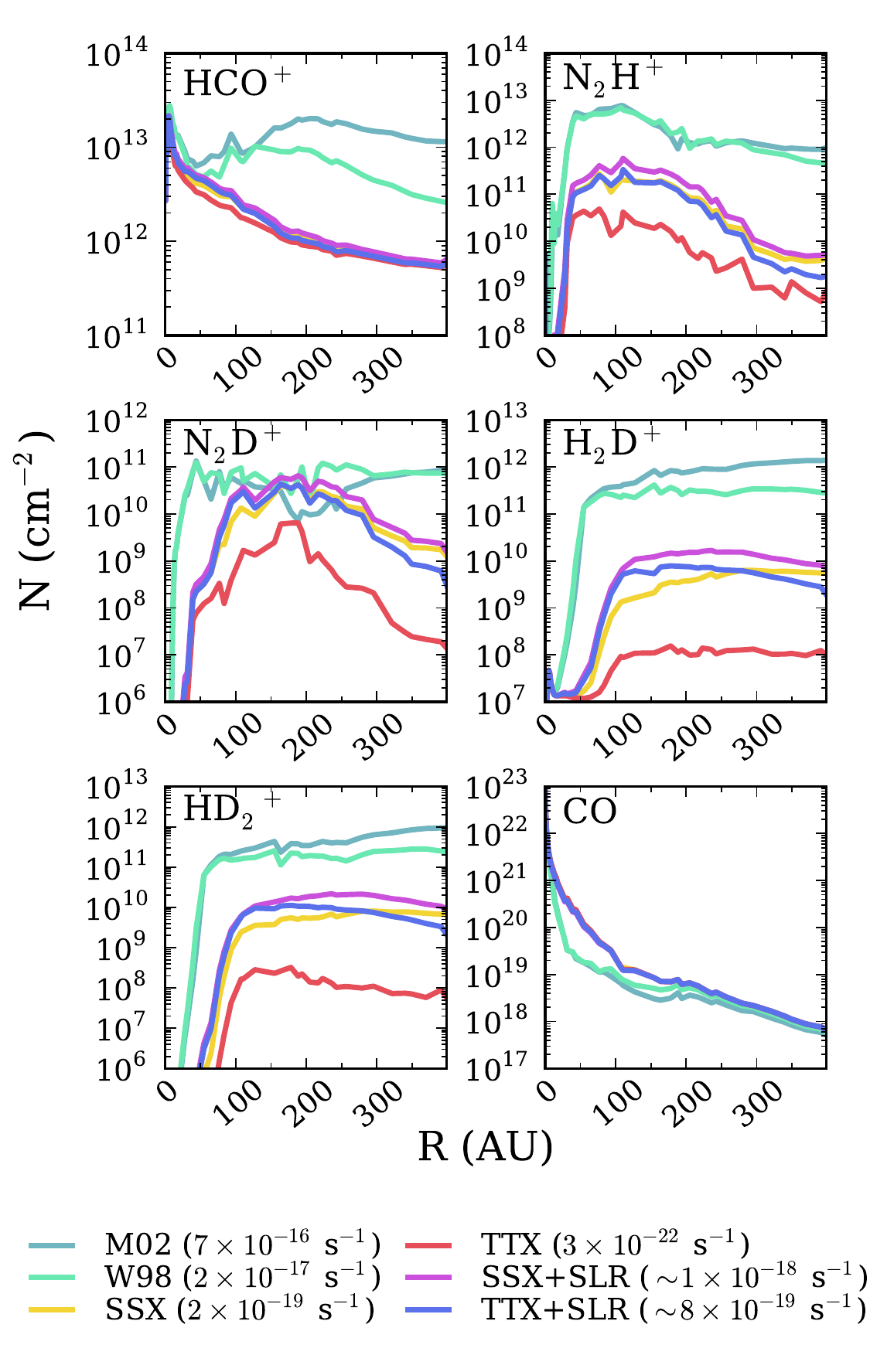}
\caption{Vertically integrated column densities (cm$^{-2}$) of the indicated species as dependent upon the magnitude of non-stellar ionization sources.  Line color indicates CR/SLR ionization model with the approximate $\zeta_{\rm{H_2}}$~s$^{-1}$ value for the incident CR ionization rates (plus SLR, if applicable), where the X-ray luminosity is fixed as $L_{\rm{XR}}=10^{29.5}$~erg~s$^{-1}$. HCO$^+$ and N$_2$H$^+$ reach a ``floor'' in their column density from a stellar X-ray ionization baseline. X-rays contribute less to N$_2$D$^+$,  H$_2$D$^+$, and HD$_2^+$ and so their column density is more sensitive to both the high and low ($\zeta_{\rm{CR}}\lesssim10^{-19}$~s$^{-1}$) ionization models.  CO column densities are provided in the lower right, and at high cosmic ray ionization rates the CO abundance is eroded by reactions with He$^+$.  \label{fig:column}}
\end{figure}

In this model framework, we compute the time-dependent chemistry as a function of radial and vertical position within the disk.  The molecular abundances for a select set of ALMA observable molecular ions and of CO after $t=1$~Myr of chemical evolution are shown in Figure~\ref{fig:abundance}.  The column headings indicate the underlying CR ionization model as described in \S\ref{sec:cr} (see also Table~\ref{tab:crrates}) with the X-ray luminosity fixed ($L_{\rm{XR}}=10^{29.5}$~erg~s$^{-1}$).  In addition to the abundances as a function of spatial position, we provide the vertically integrated column density, shown in Figure~\ref{fig:column}.  

In general, for decreasing CR ionization, the midplane ion abundances drop precipitously while the X-ray dominated surface layers are unchanged across CR models. For example, the important dense ionization tracers, \htdp\ and \ntdp, are highly abundant ($\chi\sim10^{-11}$ relative to H$_2$) in the midplane for CR ionization rates exceeding $\zeta_{\rm{CR}}\gtrsim10^{-17}$~s$^{-1}$, but drop in abundance by more than three orders of magnitude for SSX modulated rates and below.  

It is important to point out that the TTX model provides a unique chemical picture of a {\em purely X-ray dominated} disk.  In the instance of a modulated CR rate and either (i) a lack of abundant SLRs or (ii) an old ($>5$~Myr) disk in which the initial reservoir of SLRs has decayed away ($<3\%$ remaining), the scattered intermediate energy ($5-7$~keV) X-ray photons set the absolute floor to the ionization rate, with magnitude $\zeta_{\rm{XR}}\sim(1-10)\times10^{-21}$~s$^{-1}$ for a stellar X-ray luminosity of $L_{\rm{XR}}=10^{29.5}$~erg~s$^{-1}$.   In the TTX model, the midplane abundances of molecular ions are at their minimum (see especially \htdp, \hdtp, and \ntdp in Figure~\ref{fig:column}).   Correspondingly, \htdp\ and \hdtp\ have column densities typically {\em three orders of magnitude less} than if CRs are present at ISM levels and will unfortunately be difficult to detect observationally.

In the rightmost two columns of Figure~\ref{fig:abundance}, we include (static) contribution from SLR ionization, which provides an ionization rate floor of magnitude $\zeta_{\rm{SLR}}\sim(1-10)\times10^{-19}$~s$^{-1}$ for fixed initial solar nebula-like SLR abundances \citep{finocchi1997} of $^{26}$Al, $^{36}$Cl, and $^{60}$Fe.  The vertical and radial profile of the SLR ionization rate is taken from the settled disk model of \citet{cleeves2013b} (hybrid dust slab, see their Figure~2 as well as Figure~\ref{fig:ions} in the present work). For the case of SSX+SLR, the two sources contribute similarly to the total H$_2$ and helium ionization, so chemically there is little change with the inclusion of SLRs, especially in the inner disk.  The TTX+SLR model, however, increases the total ionization such that the molecular ion abundances resemble the SSX runs inside of $R\sim200$~AU. Consequently, distinguishing between a SLR and CR driven chemistry will be challenging.  Outside of $R\sim200$~AU, however, the SLR losses become important and the TTX+SLR model drops off more steeply than the SSX model, such as can be seen in \htdp.

Focusing now on the individual ions, the column density of HCO$^+$ changes by an order of magnitude between the high ($\zeta_{\rm{CR}}\gtrsim10^{-17}$~s$^{-1}$) and low CR models.  The low CR models pile up at a constant HCO$^+$ column, which physically corresponds to the minimum amount of HCO$^+$ provided by X-ray ionization in this model.  In the absence of interstellar CR, {\em the bulk of the HCO$^+$ column is therefore sensitive to only the stellar X-ray ionizing radiation} (see \S\ref{sec:xray}).  A second layer of HCO$^+$ forms at $z/r\sim0.2$, above the vertical CO freeze-out region, when CRs are present (models W98 and M02).  This layer all but disappears for the SSX and TTX models, pointing to the potential utility of disk vertical structure observations in understanding the underlying ionization environment.   It is important to note that even though we do not vary the stellar UV field in this work, for very high UV fields the CO that would otherwise form HCO$^+$ can be photo-dissociated and photo-ionized, thus forming C$^+$ in abundance.  Thus the HCO$^+$ will be indirectly UV sensitive, particularly for UV luminous Herbig Ae/Be stars \citep{jonkheid2007}, which can be orders of magnitude brighter than T Tauri stars, or in the case of extremely externally irradiated disks.  Nonetheless, the general trends over different ionization models should still hold, albeit with a lower overall HCO$^+$ column.  
Moreover, while the HCO$^+$ column can provide constraints on the ionization rates exceeding $\zeta_{\rm{CR}}\gtrsim10^{-17}$~s$^{-1}$, the HCO$^+$ is not very sensitive to lower CR rates due to its precursor, CO, freezing out in the CR dominated region.  Moreover, the HCO$^+$ column densities for the W98 and M02 models have a second peak in column density at $R\sim200$~AU not seen for the low CR models.  The second peak is actually a consequence of a deficit of CO at $R\sim50$~AU, seen in the same plot.  Because HCO$^+$ forms directly from CO, it is sensitive to the CO chemistry, both freeze-out and CO chemical processing as is the case for the W98 and M02 models.  

The \nthp\ column shows a three order of magnitude spread between a high CR rate (M02) and an X-ray only disk (TTX), with one order of magnitude variation between the different low CR models (SSX and TTX) and is a potential candidate as a midplane ionization tracer.  One caveat in using \nthp\ is that a tenuous ($\chi\sim10^{-11}$) surface layer of \nthp\ is sustained from the high X-ray photon flux (see Figure~\ref{fig:abundance}) even in the presence of CO, which is the major reactant of \nthp.  For more X-ray luminous stars, the surface \nthp\ may contribute more to total column and mask the midplane abundances. Alternatively, the contribution from the surface \nthp\ can be reduced if the N$_2$ binding energy is higher than 855~K, i.e., the N$_2$ value when CO and N$_2$ ice are well-mixed where the CO binding energy is the same, 855~K \citep{oberg2005}. Such is the case for an N$_2$ ice layered on an H$_2$O ice substrate \citep{collings2004}, but this surface N$_2$H$^+$ is never completely absent in any of our models.

The deuterated molecular ions considered here, \htdp, \ntdp\, and \hdtp, show a promising sensitivity to the low ionization models, spanning many orders of magnitude in their column densities.  Observations of these species will be essential for measuring midplane ionization, especially in the absence of CRs.  This result is a natural consequence of the pathways towards deuterated isotopologues being favored at low temperatures, resulting in their overabundance relative to the main isotopologue in cold ($T<50$~K) gas.  Naturally, the same gas is also the least affected by X-ray ionization, and therefore these species allow us to peer through the X-ray dominated upper layers directly to the midplane.

In addition to the ions, we plot the CO abundance and column density (Figures~\ref{fig:abundance} and \ref{fig:column}).  CO is a commonly used tracer of gas mass assuming an ISM conversion factor of CO to hydrogen mass of 10$^{-4}$ and is frequently used to determine chemical abundances per H$_2$ from an observed column of optically thin gas.  There is evidence, however, that the CO abundance may be lower, and hence gas masses from CO may be underestimated \citep{favre2013}.  A possible reason for the low observed abundance is that CO is chemically eroded over time by reactions with He$^+$.  The majority of the carbon returns back to CO eventually, but some non-zero fraction of the carbon is recycled into other carbon-bearing molecules, reducing the CO abundance over time \citep{bergin2014,furuya2014}. For the highly ionized CR models W98 and M02, the abundance of CO is visibly reduced near the inner disk midplane at around $R\sim100$~AU due to an increased abundance of He$^+$ and consequently a speed-up of the CO erosion process.  The column density plot (Figure \ref{fig:column}) also reflects this behavior in the M02 and W98 models and is an example of how ions can have a long-lasting effect even on abundant neutral species.   We emphasize that the layered structure of CO induced by the ionization-chemistry in the abundance plot (Fig.~\ref{fig:abundance}) may be smeared out in the presence of turbulent motions of the gas \citep{semenov2011,furuya2014}, which is beyond the scope of the current paper but should be explored in future work.

\section{Line Emission Modeling}
\label{sec:emit}

The chemical models demonstrate a sensitive link between abundances (and column densities) and ionization properties of the disk.  For example, the HCO$^+$ and \nthp\ column densities typically are sensitive to stellar X-rays, though not exclusively for high interstellar CR ionization rates.  The deuterated ions trace deeper gas, and probe ionization properties near the midplane, tracing ionization due to CRs and SLRs.  In this section we show how these effects are imprinted on the molecular emission and how excitation effects and opacity can ``mask'' the molecular column densities that allow us to discriminate between models.  From these emission models, we simulate realistic ALMA observations to determine the utility of emission line tracers as probes of individual ionizing agents.  

\subsection{Line Radiative Transfer}
\label{sec:lime}
For our molecular ion abundance results (\S\ref{sec:chemmod}), we have simulated observations of the ground-accessible submillimeter transitions of each species considered here.  The strength of the line emission depends on both the total column of material as well as the temperature of the emissive gas within the column.  To simplify the problem, we have simulated the emergent line intensity assuming the disk is observed face-on at a distance of $D=100$~pc.  The line radiation transfer is carried out with LIME \citep{brinch2010} where we create simulated ``perfect'' images of the line emission in 100 m~s$^{-1}$ channels in pixels of size 0.02$''$ (2~AU), which is much smaller than our desired resolution in the ALMA simulations.  For gas motions, we assume the gas is in Keplerian rotation about the star, and we include a turbulent doppler B-parameter of $v_{\rm{dop}}=100$~m~s$^{-1}$, as observations indicate that the turbulent broadening in disks is small \citep{hughes2011}.  

For the HCO$^+$ and \nthp\ emission models, we calculate the level populations in non-LTE with the collision rates of \citet{flower1999} as compiled in the Leiden LAMDA database \citep{schoier2005}.  For the latter we do not separate out the hyperfine structure lines, and hence refer to the transitions by their rotational-J states.  Given that the collisional rates for \ntdp\ are unknown, we treat the \ntdp\ level populations as in local thermodynamic equilibrium (LTE), which is a satisfactory approximation as the critical density of the ($3-2$) and ($4-3$) transitions are $n_{\rm{crit}}\sim2\times10^{5}$~cm$^{-3}$ and $\sim8\times10^{5}$~cm$^{-3}$, respectively, as estimated from the \citet{flower1999} collision rates for HCO$^+$ and line parameters (Einstein A-coefficients, frequencies, statistical weights) from the JPL Database\footnote{http://spec.jpl.nasa.gov/} \citep{pickett1998}, originally measured in \citet{anderson1977,sastry1981}.  We note that \citet{hugo2009} provide inelastic collisional rates for the excitation of \htdp; however, to simplify the calculations we assume the \htdp\ level populations are also in LTE.  This assumption is appropriate as the \htdp\ emission originates from dense gas exceeding n$_{\rm{H_2}}>10^{8}$~cm$^{-3}$, much larger than the critical density for  o-\htdp\ ($1_{10}-1_{11}$), $n_{\rm{crit}}\sim10^{5}$~cm$^{-3}$ \citep{hugo2009}.  

For the carbon and oxygen isotopes, we adopt ${\rm^{16}O/^{18}O=500}$ \citep{kahane1992,prantzos1996} and ${\rm^{12}C/^{13}C=60}$ \citep{keene1998}.  The deuterium isotopologues of the molecular ions are calculated within the code where the main gas reservoir (molecular hydrogen) has initially $\rm D/H=2\times10^{-5}$ (see \S\ref{sec:chemmod}). 
Leiden LAMDA formatted \htdp\ and \ntdp\ input files were compiled from the CDMS database\footnote{http://www.astro.uni-koeln.de/cdms/catalog} \citep{muller2001,muller2005} and from the JPL database, where the primary literature regarding the line parameters can be found in \citet{saito1985,amano2005,yonezu2009} for \htdp.  All simulations are carried out using the same dust distribution and opacities from the physical structure model, where the continuum is subsequently subtracted from the resulting line emission profiles.

\subsubsection{Line Opacity}

\begin{figure*}[ht!]
\begin{centering}
\includegraphics[width=0.9\textwidth]{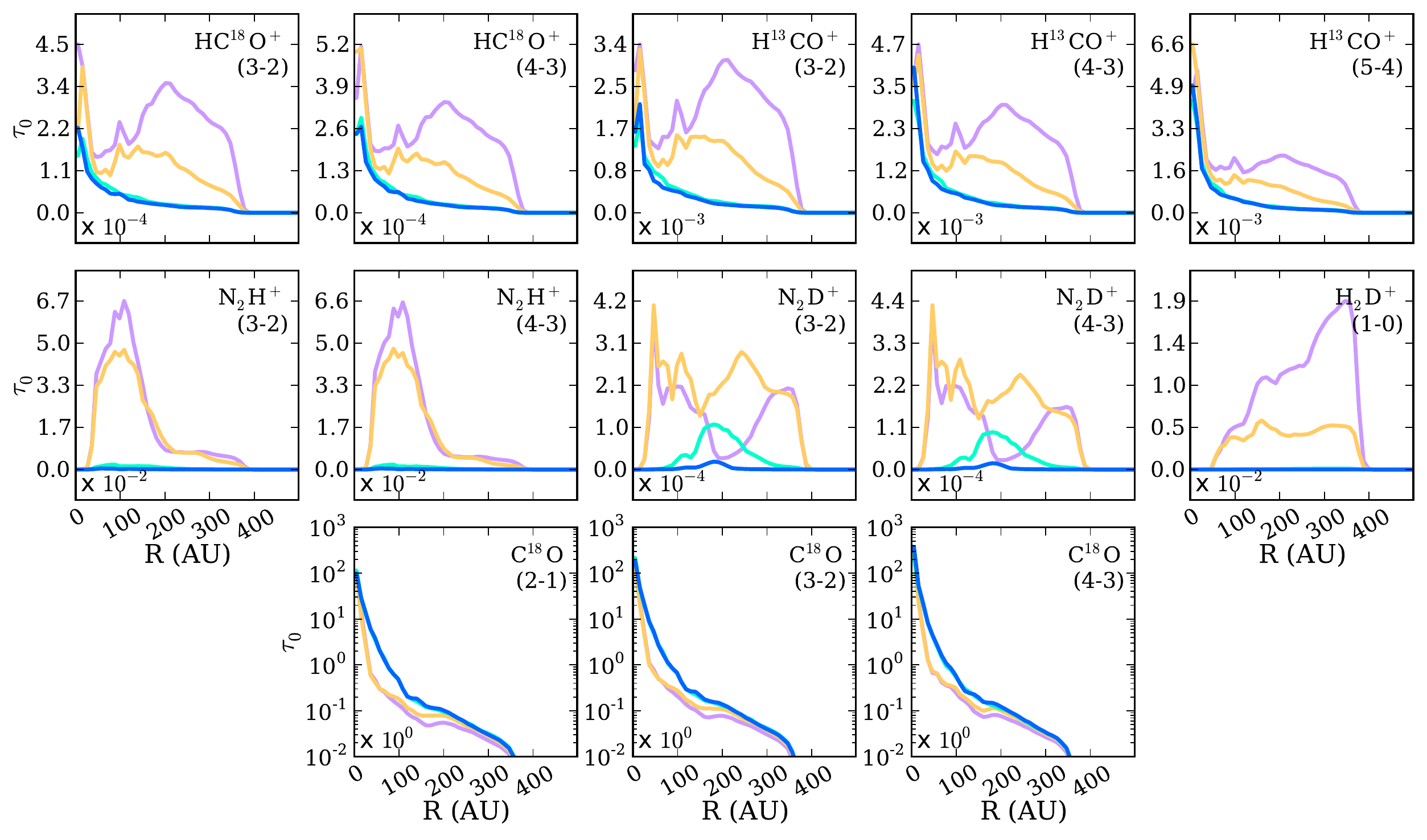}
\caption{Face-on line center optical depth ($\tau_0$) of the labeled emission lines as a function of radial distance from the star. The y-axis values are to be multiplied by the value in the lower left corner of each plot. The emission lines for the most part are optically thin except in the case of \nthp\ ($4-3$) and ($3-2$), which reaches $\tau_0$ of a few inside of $R<200$~AU.  C$^{18}$O, a commonly used tracer of gas-mass, is optically thick for most of the disk, and therefore either a more optically thin isotopologue is needed or sensitive observations of the line wings where the gas may become thin to use C$^{18}$O as a potential mass normalizer.  Colors correspond to M02: purple, W98: yellow, SSX: blue, and TTX: green. \label{fig:tau}}
\end{centering}
\end{figure*}

The LIME code is capable of providing physical line intensities, e.g., Jansky/pixel and Kelvin, as well as the line optical depth, $\tau_\nu$.  In Figure~\ref{fig:tau}, we show the line-center, vertical optical depth $\tau_0$ through the disk of the simulated transitions (direct from the emission model, i.e., no beam convolution) as a function of cylindrical disk radius.  The HCO$^+$ isotopologues are for the most part thin ($\tau_0<1$), while \nthp\ ($4-3$) and ($3-2$) reach $\tau_0\sim5$ just beyond the CO snowline for the W98 and M02 CR models. Observations of H$^{13}$CO$^+$ (3-2) and HCO$^+$ (3-2) are reported in \citet{oberg2011a} towards DM Tau.  The disk integrated flux ratio of the (3-2) transition for the $^{12}$C/$^{13}$C isotopologues is $\sim7.6$, and assuming a respective isotope ratio of 60 \citep{keene1998}, corresponds to an H$^{13}$CO$^+$ (3-2) optical depth of $\tau=0.14$.  Though this value is optically thin, it is nonetheless higher than the values from our models.  Potential explanations include a potentially higher X-ray luminosity from the star, which is unknown, or that the 19~AU inner cavity \citep{calvet2005,andrews2011} may permit the X-rays to more directly ionize the outer disk gas. 
The \ntdp\ and \htdp\ lines are thin throughout the disk for all models.  C$^{18}$O is thick, with $\tau_0\gtrsim1$ at all radii, though the CO opacity drops slightly near the star where high CR+X-ray ionization chemically destroys CO with He$^+$ more quickly than it is replenished.   We emphasize that the molecular ion opacities may be higher for higher ionization rates or if the column of gas is larger for a more massive disk, and vice versa for lower mass models (see discussion in \S\ref{sec:mass}).   

The HCO$^+$ isotopologues show a peak in their opacity offset from the radial center of the disk, where the column is highest (see Fig.~\ref{fig:column}).  The relative height between the central peak and the outer broad $100-300$~AU peak is due to a combination of column density and excitation. The ring-like feature is exaggerated for low rotational J lines such as the ($3-2$) transition where depopulation becomes important towards hotter inner disk gas, whereas H$^{13}$CO$^+$ ($5-4$) appears more centrally peaked. 

The \nthp\ lines have two main emission features, where the inner ring is caused by a contribution of stellar X-rays and CR ionization in tandem with the CO condensation front, while the outer tail at $R\sim300$ AU is fueled by primarily CR ionization.  The \htdp\ optical depth also peaks in this same outer region in the models where CR ionization is active. The \ntdp\ has a somewhat complicated $\tau_0$ profile, which is reflected in the \ntdp\ column densities (Fig~\ref{fig:column}).  The high CR ionization models sustain \ntdp\ co-spatial with \nthp\ and, when combined with warm temperatures close to the upper-state excitation temperatures ($E_u=22.2$~K and 37.0~K for $J=3$ and $J=4$ rotational levels, respectively), leads to \ntdp\ emission inside of $R<100$~AU.  

\subsection{Simulated ALMA Observations} \label{sec:simobs}
From the LIME emission models described in \S\ref{sec:lime}, we simulate full-science ALMA observations for the ground-based accessible transitions of HC$^{18}$O$^+$, H$^{13}$CO$^+$, \nthp, \ntdp, and \htdp.  The full set of emission lines considered at present are listed in Table~\ref{tab:obs}.  We note there are additional lines that are accessible by ALMA that we chose not to include, such as \ntdp\ (5-4) and \hdtp\ ($1_{10}-1_{01}$), because these lines were unobservable for all emission models considered. The full-science array is comprised of fifty 12-m antennas, twelve 7-m antennas, and four 12-m antennas, i.e., the total power (TP) array, where the 7-m and TP arrays are referred to as the Atacama Compact Array (ACA).  For each of the line models we assume the same set of simulated observing parameters, and note that in some instances there are more efficient (i.e., less ALMA time) means to achieve the same goals.  The specific settings for actual observations should be tailored to the specific target and specific line being studied.  

\begin{deluxetable}{ccc} 
\tablecolumns{4} 
\tablewidth{0pt}
\tablecaption{CASA Simulation Parameters.  \label{tab:obs}}
\tabletypesize{\footnotesize}
\tablehead{  Emission &  12-m Configuration\tablefootnote{}  & Sensitivity\tablefootnote{}  \\
 Line & (12-m + 7-m Beam) &  (mJy) }
\startdata
HC$^{18}$O$^+$ ($3-2$) & 03 ($1.28''$)  & $1.0$  \\
HC$^{18}$O$^+$ ($4-3$) & 03 ($0.96''$)  &  $1.4$  \\
H$^{13}$CO$^+$ ($3-2$) & 10 ($0.33''$) & $0.97$  \\
H$^{13}$CO$^+$ ($4-3$) & 10 ($0.25''$)  & $1.4$ \\
H$^{13}$CO$^+$ ($5-4$) & 10 ($0.20''$)  & $3.6$  \\
\nthp\ ($3-2$)                     & 10 ($0.31''$)  & $1.3$  \\
\nthp\ ($4-3$)                     & 10 ($0.24''$)& $2.6$  \\
\ntdp\ ($3-2$)                     & 01 ($1.55''$) &$1.3$  \\
\ntdp\ ($4-3$)                     & 01 ($1.16''$) &$1.3$  \\
o-\htdp $\left(1_{10}-1_{11}\right)$     & 01 ($0.96''$)  &$2.5$  
\enddata
\tablenotetext{a}{http://casaguides.nrao.edu/index.php?title$=$Antenna\_List  Beam averaged over major and minor axes from the combined 12-m and ACA observations.}. 
\tablenotetext{b}{6.1$h$ in 0.2 km/s bins from the ALMA Sensitivity Calculator.} 
\end{deluxetable}

\subsubsection{Observational Parameters}
The simulations presented here reflect 6$h$ of total on-sky time with one of the 12-m array configurations and an ACA 7-m observation.  While the choice of 6$h$ is somewhat arbitrary, it represents a relatively ``deep'' observation, and ideally one would target more than one emission line towards a given source in the same proposal. Thus this  length of time may be somewhat optimistic, but is designed to detect the lines with high signal-to-noise.  Some of the lines are undetectable for low ionization models as is the case for \htdp, and thus deeper observations would be needed to detect the line if that specific model reflects the true properties of the disk.  In that case, going to a different tracer such as \ntdp\ may be a better choice than a deeper observation of a weak line.  Furthermore, these results can be approximately scaled for sources at different distances and the estimated signal-to-noise can be adjusted for different observation times.

Regarding antenna configurations, for stronger emission lines, such as H$^{13}$CO$^+$, and \nthp, we choose somewhat more extended 12-m configurations (see Table~\ref{tab:obs}), while for weaker lines, we simulate the most extended configuration that still provides high signal-to-noise if possible, which in some instances is the most compact, least-resolved full-operations configuration, denoted 01.  We include ACA 7-m observations because the maximum scale of the $R=400$~AU disk at $d=100$~pc on the sky is 8$''$, and for the higher frequency ($\nu>300$~GHz) lines considered here, the 7-m array recovers the flux at all scales without the need for the TP array.  At lower frequencies, a second 12-m configuration can be used instead to recover all of the flux in significantly less time than for the 7-m observations, which is clearly the better -- and default -- option for on-sky ALMA observations.  Additionally, the long duration of the simulated observations fills out the UV coverage and, due to sky-projection effects, can decrease the minimum baseline, allowing for slightly larger maximum recoverable scales. 

\begin{figure}[ht!]
\begin{center}
\includegraphics[width=0.48\textwidth]{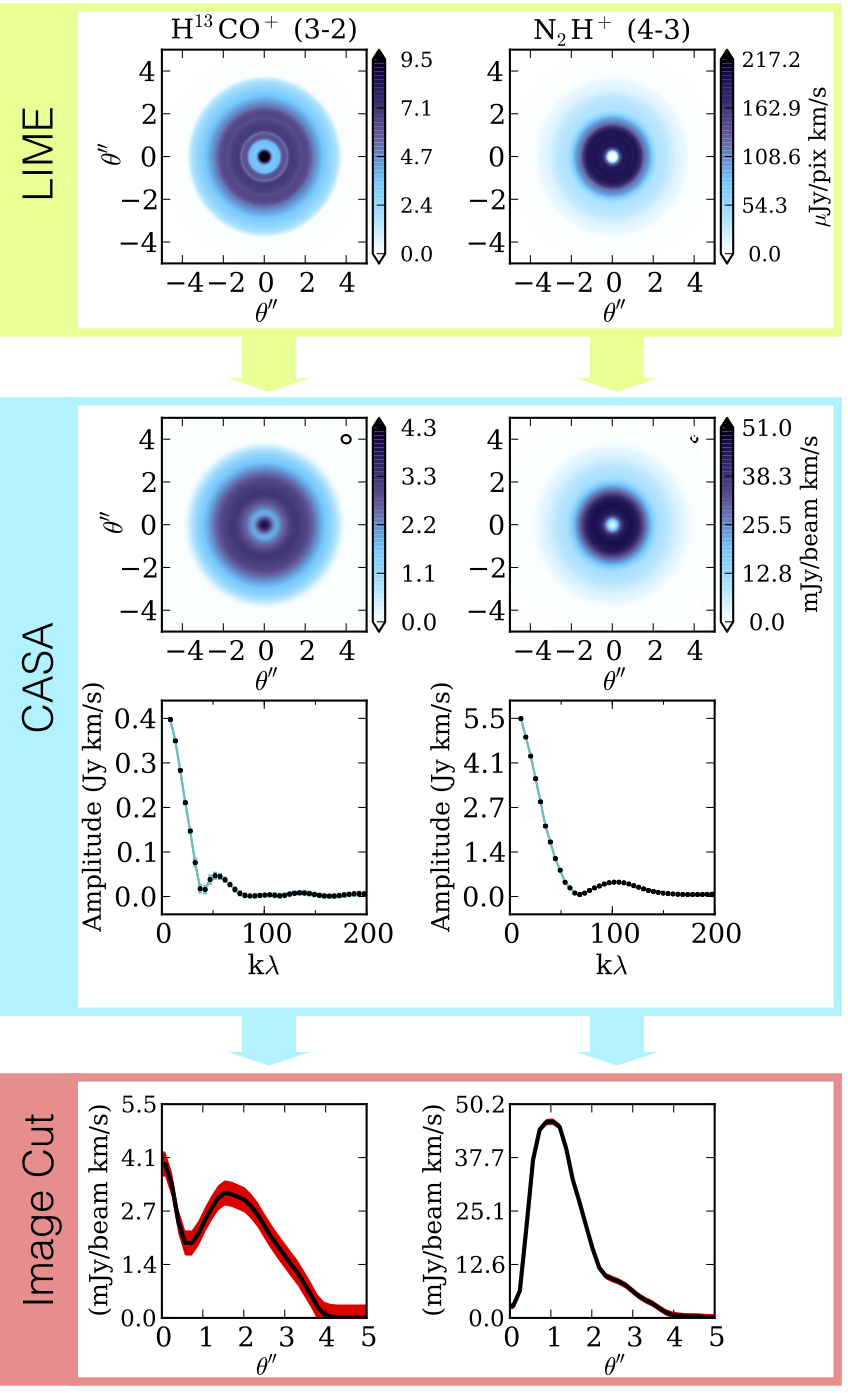}
\caption{Schematic illustrating the calculation of simulated observations. {\em Top:} The face-on line emission models calculated from the chemical abundance models (Fig.~\ref{fig:abundance}) using the radiative transfer code LIME \citep{brinch2010} in non-LTE where available.  Line emission models are in units of $\mu$Jy/pixel and are observed at a distance of $d=100$~pc.  {\em Middle:} Using CASA's simobserve task, we simulate observations with the full 12m ALMA array plus the 7m ACA array.  UV-binned visibility amplitudes integrated over velocity and on-sky emission shown, $1''=100$~AU.  {\em Bottom:} Cuts across the on-sky emission from the cleaned simulated observations, see text for details.   Error bars (sensitivity) are estimated from the ALMA sensitivity calculator values assuming a 6.1$h$ observation in 100 km/s channels.  Average beam size is $0.33''$ and $0.24''$ for the left and right plots, respectively.
\label{fig:flow}}
\end{center}
\end{figure}

In summary, the simulations presented here are designed to show the sensitivity of the emission line observations to chemical signatures of ionization processes in disks. These results provide a primer to aid in quantifying the contribution of different physical processes to disk ionization, a crucial ingredient in models of disk chemistry. We note that the specific choice of observing parameters will depend upon properties of the source, the ALMA cycle, which determines the available antenna configurations, and the lines targeted.  Figure~\ref{fig:flow} illustrates the full process beginning with LIME emission models to creating ALMA on-sky simulations for two example emission lines.
Rather than showing images or channel maps for every emission line/model considered both in model emission (\S\ref{sec:lime}) and CASA simulations, we instead present our results as circularly averaged visibility amplitudes and circularly averaged integrated line intensity cuts in sky coordinates, which facilitates easy comparison between observations of different ionization models for a given emission line.

\subsubsection{ALMA Sensitivity Calculation}\label{sec:sensitivity}
To calculate the uncertainty on the ALMA measurements, we use the uncertainties from the ALMA sensitivity calculator\footnote{https://almascience.nrao.edu/proposing/sensitivity-calculator}, which set the width of the shaded regions in the image plots (rightmost column of Figures~\ref{fig:almahcop} and \ref{fig:alman2h}). We do not simulate thermal noise within the CASA simulations themselves to save computational time. To estimate the error on individual baselines at each time sampling (defined by integration time, which was $t_{\rm{int}}=1000$s  for the simulations), we scale the sensitivity calculator's uncertainty $\sigma_0$ by $\sigma_i=\sigma_0 \sqrt{N_A(N_A-1)} \sqrt{N_{\rm{time}}}\sqrt{N_{\rm{pol}}}$, where $N_A$ is the number of antennas (50 for the 12-m array and 12 for the 7-m array), $N_{\rm{time}}$ is the number of time samplings during the course of the simulated observation's integration time (e.g., $t_{\rm{obs}}=22000s$ $=6.1h$ on the 12-m array such that $N_{\rm{time}}$=22) and $N_{\rm{pol}}=2$ is the number of polarizations, where the emission is unpolarized. We note that if we had chosen a smaller (larger) sampling time within a factor of a few, we would have correspondingly more (less) points to bin over, and the final error on the binned measurement is not very sensitive to the choice of sampling time. 
From the full measurement set, we bin the velocity-integrated visibility amplitudes by projected UV distance.  The uncertainty on the binned value (the mean) is taken to be the error on the individual measurements divided by the number of points within the UV distance bin, $\sigma_A=\sigma_i/\sqrt{N_{\rm bin}}$, where $N_{\rm bin}$ is typically $\gtrsim100$. 

\subsubsection{Emission Model Results}
\begin{figure*}[ht!]
\centering
\includegraphics[width=0.75\textwidth]{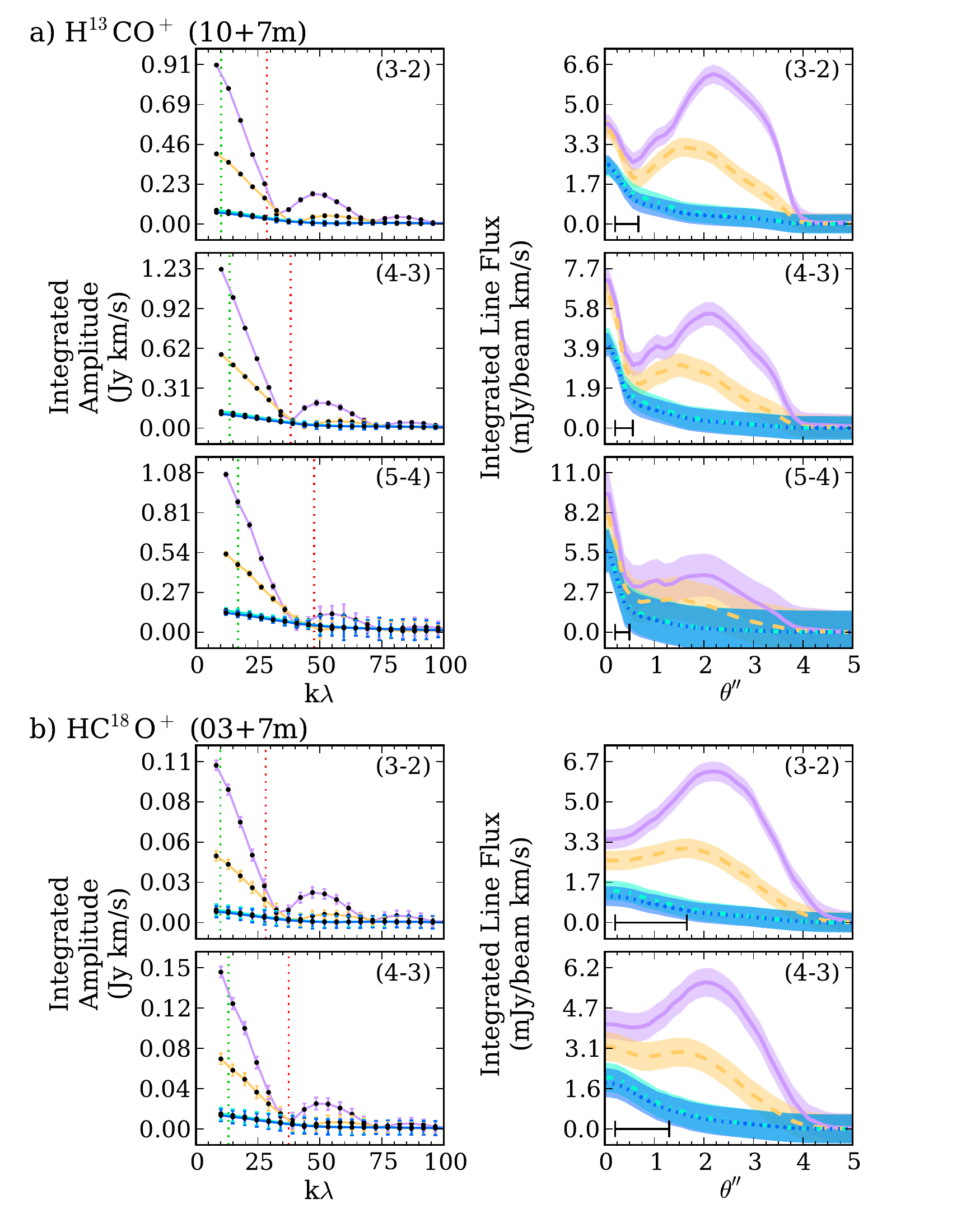} 
\caption{ALMA simulations for the observable transitions of the indicated molecular ions.  Line color represents different CR ionization models: M02 (purple), W98 (yellow),  SSX (cyan), and TTX (blue). The simulated antenna configuration used in CASA's simobserve task is indicated in parenthesis (\#\# + 7m).  {\em Left:} Velocity-integrated visibility amplitudes.  The dashed vertical green line corresponds to the minimum baseline from the 12-m observations, which can be less than the physical separation of the antennas due to sky projection. The red dashed line is the maximum baseline sampled by the 7-m observations.  {\em Right:} reconstructed image profiles in sky-coordinates integrated over velocity.   Synthesized beam indicated by the black bar in the bottom left corner of the image plots.  The width of the individual lines corresponds to the sensitivity of the simulated observations, see \S\ref{sec:sensitivity} for details. \label{fig:almahcop}}
\end{figure*}

\begin{figure*}[ht!]
\centering
\includegraphics[width=0.75\textwidth]{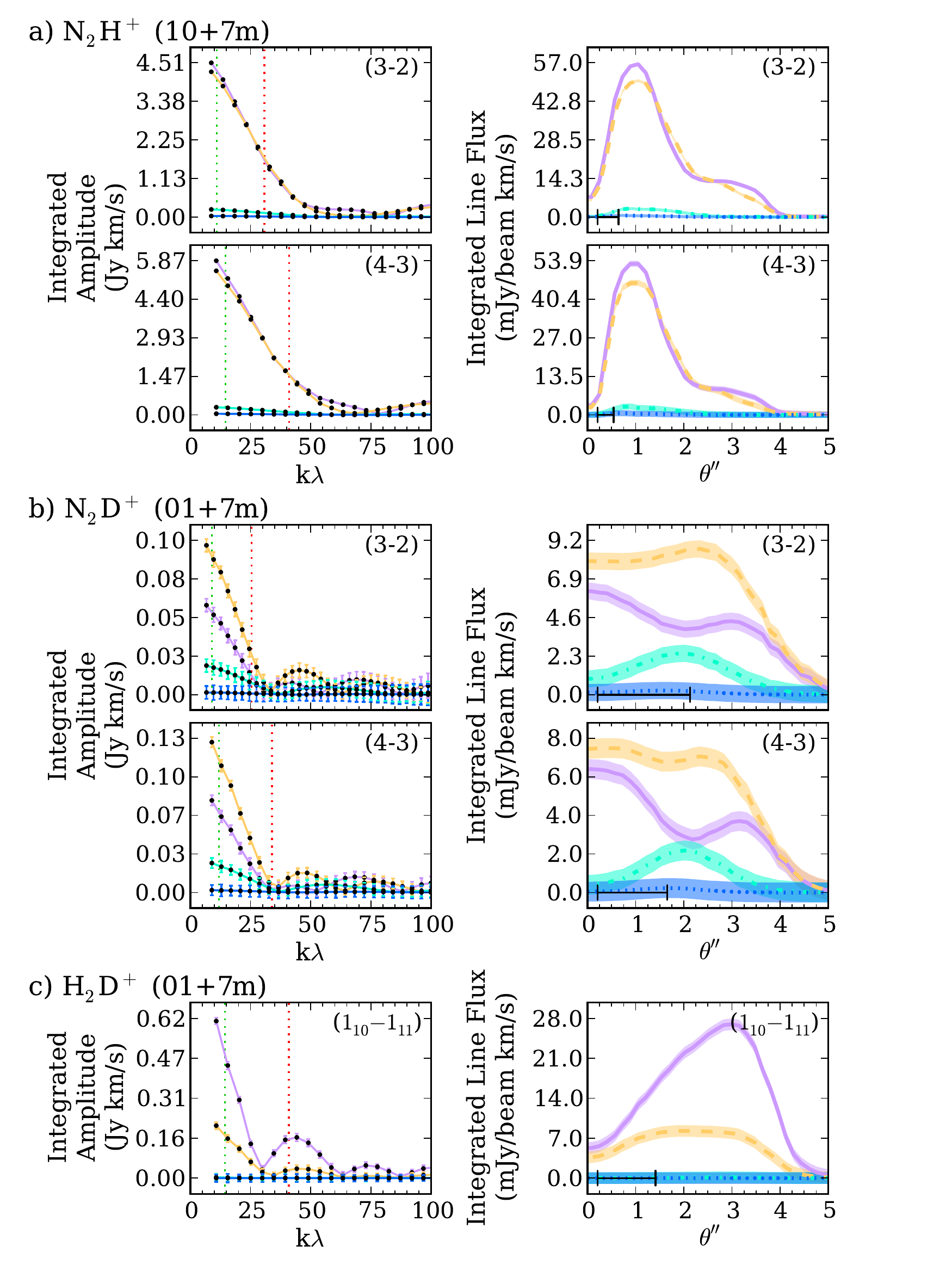}
\caption{The same as Fig.~\ref{fig:almahcop}, for ``cold gas'' ions typically tracing gas below $T\lesssim50$~K.  H$_2$D$^+$ is unobservable for our models with ionization rates of $\zeta_{\rm{CR}}\lesssim2\times10^{-18}$~s$^{-1}$. Consequently, N$_2$D$^+$ is the only molecular ion which allows for some differentiation between ionization rates below $\zeta_{\rm{CR}}\lesssim10^{-18}$~s$^{-1}$, as seen clearly in the visibilities, especially for the \ntdp\ (4-3) transition.  The high CR ionization rate models overlap for \nthp\ because the line becomes very optically thick.  We note, that while weak, the SSX model is detectable in both the ($4-3$) and ($3-2$) transitions of \nthp\ at the 3 and 6$\sigma$ levels, respectively.  See also Figure~\ref{fig:almaslr} for a zoom-in on the low ionization models.\label{fig:alman2h}}
\end{figure*}

We present the primary results of the ALMA simulated observations in Figures~\ref{fig:almahcop} and \ref{fig:alman2h}. The CR ionization models described in \S\ref{sec:cr} are indicated by line color, where purple represents the highest, diffuse ISM CR rate, while the blue curve is representative of an X-ray dominated chemistry.  
The identifier of the full-operations array used to simulate each set of data is indicated in parenthesis next to the name of the species (see Table~\ref{tab:obs} for equivalent angular resolution). The left and right columns are different representations of the same simulated data.  The left column shows the visibility amplitudes integrated over the line (Jy~km~s$^{-1}$) where the quantity ``integrated amplitude'' corresponds to the fact that we have integrated over the channels that contain line flux rather than averaging, which is often done for continuum amplitudes.  The visibility amplitude plots are labeled with a dashed green vertical line indicating the maximum recoverable scale for the 12-m array observations alone.  The dashed red vertical line indicates the minimum scale sampled by the 7-m array.  We note that the minimum baseline indicated by the green line may be less than what would be predicted for the same array based upon the physical baselines due to sky-projection effects mentioned previously. The right column shows the brightness profile of the reconstructed (cleaned) images on the sky, where $\theta$ is relative to the position of the central star and $1''=100$~AU. To create the profiles, we average ten radial slices across the face of the disk.  The noise reflects the sensitivity on a single cut, but in practice averaging over slices can further reduce the noise on the profile.  

The HCO$^+$ isotopologues' line emission (Figure~\ref{fig:almahcop}) is sensitive to both the stellar X-rays and the high CR ionization models (M02 and W98 in purple and yellow respectively).  This behavior is reflected in the column densities (Figure~\ref{fig:column}) and in the simulated observations, both in the visibilities and reconstructed images.   The line intensity is indistinguishable for lower CR rates for models TTX and SSX, green and blue, where stellar X-rays -- and not CRs -- set the minimum HCO$^+$ abundance in this particular model.  The high M02 CR rates form a clear emission ring (i.e. the HCO$^+$ emission peaks offset from the star) for the ($3-2$) transition, less prominently seen in the ($4-3$).  The reason that the ($3-2$) emission is brightest away from the column density peak is due to excitation of the line, which has an upper state energy of $E_u\sim25$~K.  The abundant HCO$^+$ gas inside of $R\lesssim50$~AU has temperatures typically exceeding 100~K.  At this point, higher J-states of H$^{13}$CO$^+$ and HC$^{18}$O$^+$ become more populated, thereby de-populating the J~=~3 levels and decreasing the column of emissive gas, behavior reflected in the line optical depths (Figure~\ref{fig:tau}) and emission plots (Figure~\ref{fig:almahcop}).   
For higher rotational transitions of H$^{13}$CO$^+$, ($4-3$) and ($5-4$), there is an inner ($\theta<0.5''$) peak in the simulated observations (Figure~\ref{fig:almahcop}) due to a confluence of high X-ray rates and abundant gas-phase CO.  This feature is also present in the underlying HC$^{18}$O$^+$ abundances, but the lower spatial resolution simulations adopted for HC$^{18}$O$^+$ (see Table~\ref{tab:obs}) do not resolve the inner 50~AU.  In summary, the HCO$^+$ emission sensitively traces i) stellar X-ray processes and ii) high CR fluxes, but is limited in its capacity as a midplane tracer due to CO freeze-out in the outer disk. Furthermore, the low-J HCO$^+$ behavior highlights a situation in which observations of emission rings are {\em excitation effects,} rather than chemical or physical structure, such as a planet or a snowline.  

\begin{figure*}[ht!]
\centering
\includegraphics[width=0.75\textwidth]{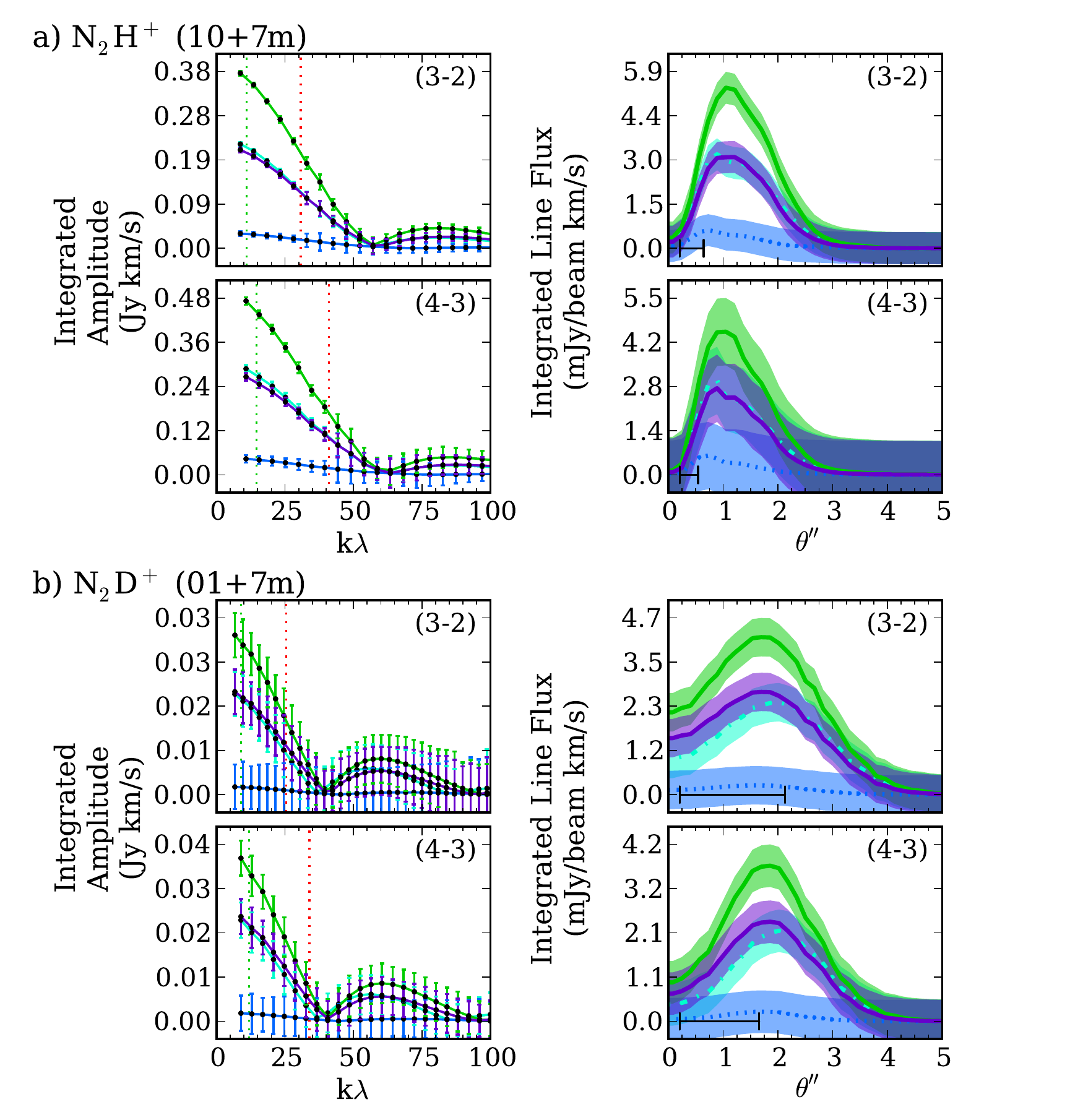}
\caption{ALMA simulations of the low CR ionization rate models with and without a constant SLR ionization rate.  Only the low ionization models are shown for clarity.  Figure quantities are the same as Figure~\ref{fig:almahcop}.  The CR-only models are shown as TTX: blue and SSX: cyan.  The same CR models now including SLR ionization are shown in green and purple, respectively. As can be seen from both visibility curves and sky-emission, the inclusion of SLR ionization does not significantly change the profile or intensity of the lines shown.  A TTX model with SLR ionization (purple) would be difficult to distinguish from an SSX model without SLRs (cyan). \label{fig:almaslr}}
\end{figure*}

 \nthp\ is expected to arise from cold gas where its destroyer, CO, is frozen out at temperatures below $T<20$~K \citep{aikawa2001,bergin2002,jorgensen2004}.  Nevertheless, in the strongly irradiated X-ray layers we find that some \nthp\ is sustained even in the presence of CO.  Consequently, its strength as a diagnostic tool of midplane ionization is somewhat decreased by potential confusion with surface \nthp\ emission.  As shown in Figure~\ref{fig:alman2h}, the high ionization and low ionization models are clearly discernible, while the SSX and TTX models are -- in a relative sense -- far more difficult to tell apart.  We note \nthp\ may nonetheless have utility as a midplane ionization tracer, and that the SSX and TTX models are both observable and differ by a factor of $\sim7$ in brightness (see the zoom-in in Figure~\ref{fig:almaslr}), but detailed modeling may be required to estimate the potential X-ray contribution to the \nthp\ column density.  An additional caveat in using \nthp\ as an ionization tracer is that its emission can be partially optically thick (if not completely, $\tau\sim5$; Fig.~\ref{fig:tau}).  Indeed, the high interstellar CR models, M02 and W98, have similar emission line strengths because the column densities are quite large, and the correspondingly thick emission lines are no longer sensitive to column density.  Therefore, in this model, \nthp\ observations would help distinguish between disks where CRs are present or where they are excluded, e.g., by winds, however it is perhaps not a precise tool for determining CR rates below $\zeta_{\rm CR}\sim10^{-19}$~s$^{-1}$ for this particular disk structure. In Figure~\ref{fig:alman2h}, we note that even though the SSX model emission is substantially lower than that of the W98 and M02 models, it nonetheless is still detectable.  The width of the radial profile corresponds to the ALMA sensitivity, and thus disks with SSX-level of CR ionization are detectable at the $3\sigma$ and $6\sigma$ level for the ($4-3$) and ($3-2$) transitions of \nthp, respectively.  Given the \nthp abundance's concurrent chemical dependence on temperature, a warmer disk where the \nthp\ abundance peaks further from the X-ray bright star may allow \nthp\ to be a more sensitive tracer of non-stellar ionizing processes and be less optically thick at the emission peak due to lower outer disk gas densities.

\ntdp\ shows interesting emission behavior such that the M02 models show {\em less} \ntdp\ emission than the more weakly CR-ionized W98 models (Figure~\ref{fig:alman2h}).  This behavior is reflected in the column densities (Fig.~\ref{fig:column}) and the abundances (Fig.~\ref{fig:abundance}). More specifically, the M02 \ntdp\ models show a deficit in abundance relative to the W98 models centered at $R\sim200$~AU.  The inner \ntdp\ gas is co-spatial with the high \nthp\ (fueled by X-ray flux) and with the boundary of the \htdp\ abundant region ($T<50$~K), creating an \ntdp\ layer.   This layer is reflected in the W98 models as well.  At intermediate distance radii $R\sim150-300$~AU, the M02 models show an \ntdp\ deficit as a direct consequence of the high degree of CR ionization in the presence of cold gas where freeze-out becomes important.  The combination of ionization and cold gas drives important non-equilibrium chemistry.  More specifically, the nitrogen is not being recycled back into \nthp\ but instead sequestered into nitrogen bearing ices like NH$_3$, HCN and NO, and therefore is not available as N$_2$ in the gas to reform \nthp\ and \ntdp.  This process is an analogue to a similar sequence that may occur for CO \citep{bergin2014}.  In the M02 model, these pathways are specifically triggered by M02's high CR flux, but a factor of $\sim10$ brighter X-ray luminosity would have a similar effect (see \S\ref{sec:xray}).

The true utility of \ntdp\ is highlighted at low CR ionization rates. The SSX models are clearly distinguishable from the TTX models, which is reflected in the emission plane for both ($3-2$) and ($4-3$) and in the column density plane.  \ntdp\ emission is always very optically thin, making it a direct tracer of column regardless of CR rate.  The difference between the SSX and TTX line intensities is a factor of about $\sim20$, more than double that of \nthp, allowing these CR models to be more easily disentangled than with \nthp.

Finally, \htdp\ is a commonly used cold ionization tracer.  However, as can be seen in Figure~\ref{fig:alman2h}, its emission is only detectable for interstellar CR rates or higher, owing to its weak line strength.  For ionization rates at or below $\zeta_{\rm{CR}}\lesssim10^{-19}$~s$^{-1}$, the \htdp\ ($1_{10}-1_{11}$) is undetectable even for full ALMA operations, and therefore it is only a useful tracer of interstellar CR rates, if they are present.  These results are consistent with existing limits on the observed \htdp\ column towards the TW Hya protoplanetary disk \citep{chapillon2011,qi2008}; however, such limits are much higher than all of the line strengths predicted here and thus more sensitive observations are required to determine if CRs are present with \htdp\ as a tracer.  The utility of \htdp\ is highlighted when used in tandem with \ntdp.  As mentioned above, \ntdp\ decreases in brightness for both high ionization rates (when the precursors of \ntdp\ are chemically destroyed) and low ionization rates (when \ntdp\ is not produced).  Observations of \htdp\ in conjunction with \ntdp\ would allow one to break the degeneracy between these scenarios.

In Figure~\ref{fig:almaslr}, we show simulated ALMA observations for the models including SLRs (no time decay) for the SSX and TTX models.  The \nthp\ and \ntdp\ lines are the only tracers for which there may be a measurable difference for the low ionization models.  From these plots it is apparent that distinguishing between CR fueled chemistry and SLR chemistry is extremely difficult.   Additional factors such as -- highly uncertain -- disk ages will be necessary to determine fractional contributions.  Alternatively, if an otherwise unexpected ``jump'' in ionization is seen, e.g., at the boundary of a T-Tauriosphere, then the contribution from each component can be determined unambiguously.  Without this additional information, however, measurements of dense gas ionization using emission line tracers will most likely reflect a combination of both CR and SLR effects.

\section{Further Considerations}\label{sec:fc}
In this section, we relax certain assumptions of our model and explore how our model results depend upon these additional parameters, including X-ray luminosity, temporal decay of SLRs, and the assumed mass of the disk. 
\subsection{Higher X-ray Luminosity}\label{sec:xray}
\begin{figure*}[ht!]
\centering
\includegraphics[width=0.8\textwidth]{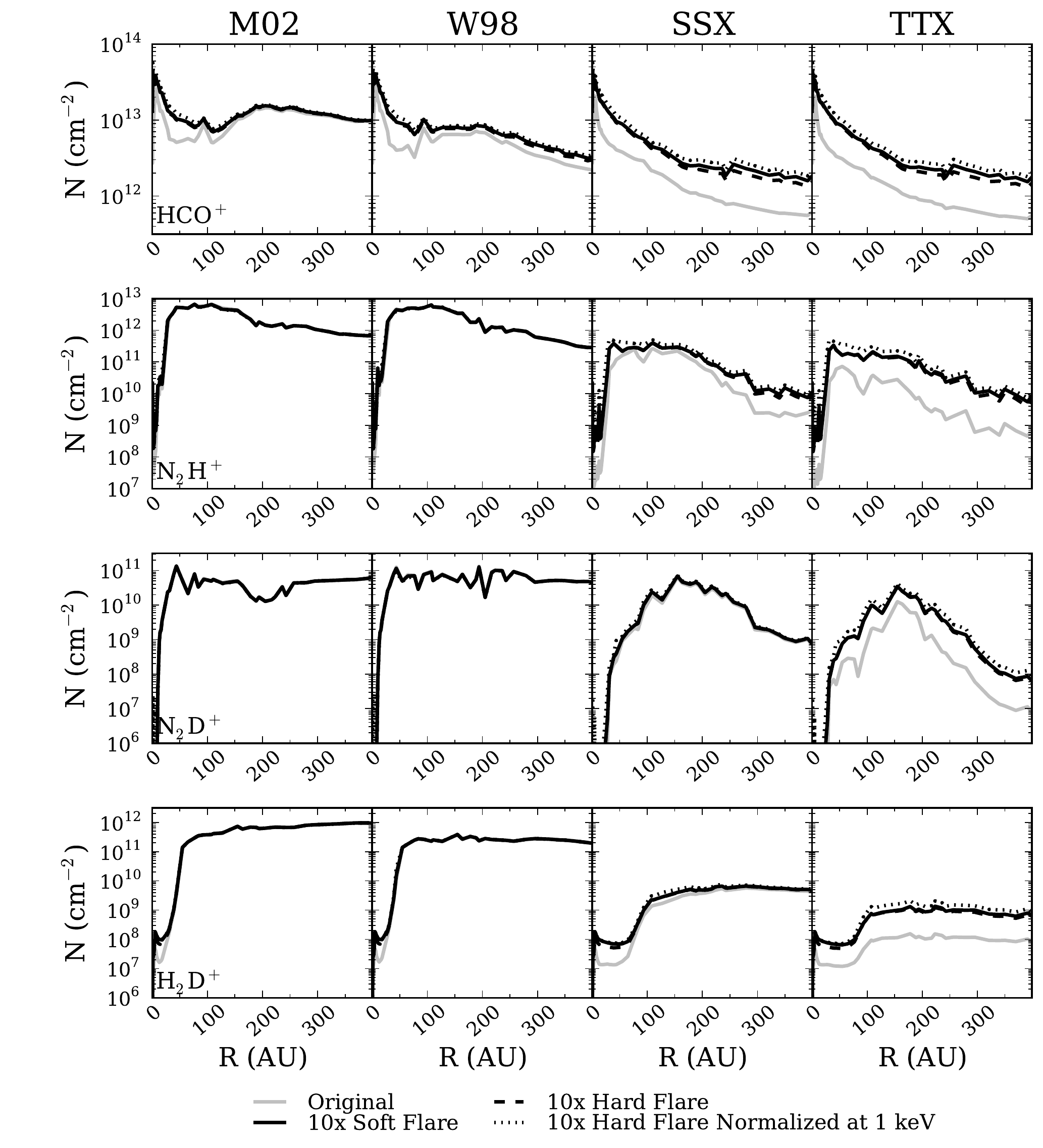}
\caption{Chemical models for an X-ray rate of $L_{\rm{XR}}=10^{29.5}$~erg~s$^{-1}$ (gray solid lines) and an enhanced rate of $L_{\rm{XR}}=10^{30.5}$~erg~s$^{-1}$ (black lines).  Rows correspond to the indicated emission lines as labeled in the left column, and columns correspond to the CR models as labeled at the top of the figure.  For the enhanced models, the three line styles correspond to different X-ray spectral templates.  The solid black line holds the spectral shape fixed and increases the overall luminosity by a factor of 10. The dashed black line corresponds to a harder X-ray spectrum with the same luminosity as the solid line, normalized such that there are fewer photons with $E_{\rm XR}<3$~keV and more with $E_{\rm XR}>3$. The dotted line is a model with the same 1 keV flux as the solid black line, but with a hard spectral template such that the X-ray luminosity is also slightly higher, a factor of 1.7, due to the hard X-ray tail). \label{fig:xrays}}
\end{figure*}
The results presented in Figures~\ref{fig:almahcop}--\ref{fig:almaslr} consider a single X-ray luminosity, $L_{\rm{XR}}=10^{29.5}$~erg~s$^{-1}$.  To understand the sensitivity of the lines to X-ray ionization, we have computed an additional chemical model for a ten-fold increase in X-ray luminosity  for three spectral shapes: (1) the same ``quiescent'' spectral shape for the baseline model, solid line; (2) a harder X-ray spectrum as was used in \citet{cleeves2013a}, with the same normalized luminosity as model (1), dashed line; and (3) the hard X-ray spectrum normalized to have the same X-ray flux at 1 keV as model (1), dotted line.   The results of these higher X-ray luminosity models are shown in Figure~\ref{fig:xrays}, where the standard model is shown as the gray solid line and the elevated models are shown in black.   

The high X-ray ionization rate changes the HCO$^+$ column density most significantly in the inner $R<100$~AU for the M02 and W98 models; for the outer disk there is a reduced effect as the contribution by CRs is more important.   For the reduced CR models, the HCO$^+$ column density is enhanced throughout the disk by a factor of 3-4.
This trend is explained by the balance between ionization and recombination, such that the steady state abundance of ions (or electrons) is proportional to $\propto\sqrt{\zeta/(\alpha n_{\rm{H_2}})}$, where $\alpha$ is the recombination rate and $n_{\rm{H_2}}$ is the number density of H$_2$. With everything else constant, a factor of ten increase in ionization results in a factor of $\sqrt{10}\approx3$ increase in the abundance and consequently column density.  Consequently, owing to the sensitivity of HCO$^+$ to the X-ray ionization field, observations of optically thin isotopologues of HCO$^+$ may help put constraints on the permeability (optical depth) of the disk gas and dust to X-ray photoionization if the stellar X-ray luminosity is known.  Furthermore, additional constraints on the disk gas mass combined with X-ray measurements would provide an approximate measure of the opacity due to dust and gas in the X-ray irradiated layers.  Variations induced by the difference in spectral templates (black lines) are smaller, typically a factor of 1.2-1.8 in the TTX models.  The hard X-ray spectrum considered in (2) (dashed line) even has a slightly lower molecular ion column density compared to the softer X-ray spectrum of (1) owing to the initial destruction of CO and N$_2$ by hard X-ray generated He$^+$.  Model (3), i.e., the spectrum normalized at 1 keV with model (1), has a slightly higher energy integrated luminosity, 1.7~times that of models (1) and (2), at which point production overtakes the destruction of precursors. The \nthp, \ntdp, and \htdp\ column densities do not become sensitive to variations in the X-ray luminosity and spectrum until very low CR rates, primarily TTX, where most of the ionization in these cases comes from stellar X-rays.  The flat increase in ionization for the TTX models is a consequence of high X-ray ionization rates near the star and slow recombination at large radii, where the drop in density (recombination) balances the decrease in X-ray flux with distance from the star.  

\subsection{Short-Lived Radionuclide Time Decay}\label{sec:rntime}
Another simplifying assumption made in the chemical abundance calculations in Figures~\ref{fig:abundance} and \ref{fig:column} was the use of a constant, non-decaying SLR ionization rate for the models including radionuclide ionization sources.  From the results of \citet{cleeves2013b} (see also Appendix A), the ensemble of short-lived radionuclides that dominate the ionization, namely $^{26}$Al and $^{60}$Fe, have an effective net half-life of approximately $t_{\rm half}\sim1.2$~Myr (see \S\ref{sec:slr} for further details).  For disk lifetimes up to 1~Myr, the change in ionization rate is correspondingly less than a factor of two, but can become significant for older disks ($>3$~Myr). To simulate the decay of ionizing SLRs with time within the disk chemistry code, we have created a model where the SLR ionization rate is now time-dependent internal to the chemical calculations and the total (ensemble) rate decays with a 1.2~Myr half-life (Appendix~A), as a simple first order approximation.  In reality, the specific radionuclide decay products (i.e., $\gamma$-rays, $\beta^+$ particles, etc.) will evolve as each parent nuclide decays, i.e., $^{26}$Al versus $^{60}$Fe and will result in different amounts of energy deposited or lost depending upon the ionization cross sections and the decay rate of the parent nuclide.  The changeover from a $^{26}$Al dominated SLR rate to $^{60}$Fe  rate happens at around 5~Myr, and so strictly speaking a more detailed treatment that is beyond the scope of this paper would separate the individual contributions.  We note that in addition to time variation, there will be variation between the starting abundances of SLRs between disks.  In this analysis we assume solar nebula-like abundances from \citet{cleeves2013b}.

\begin{figure}[ht!]
\centering
\includegraphics[width=0.49\textwidth]{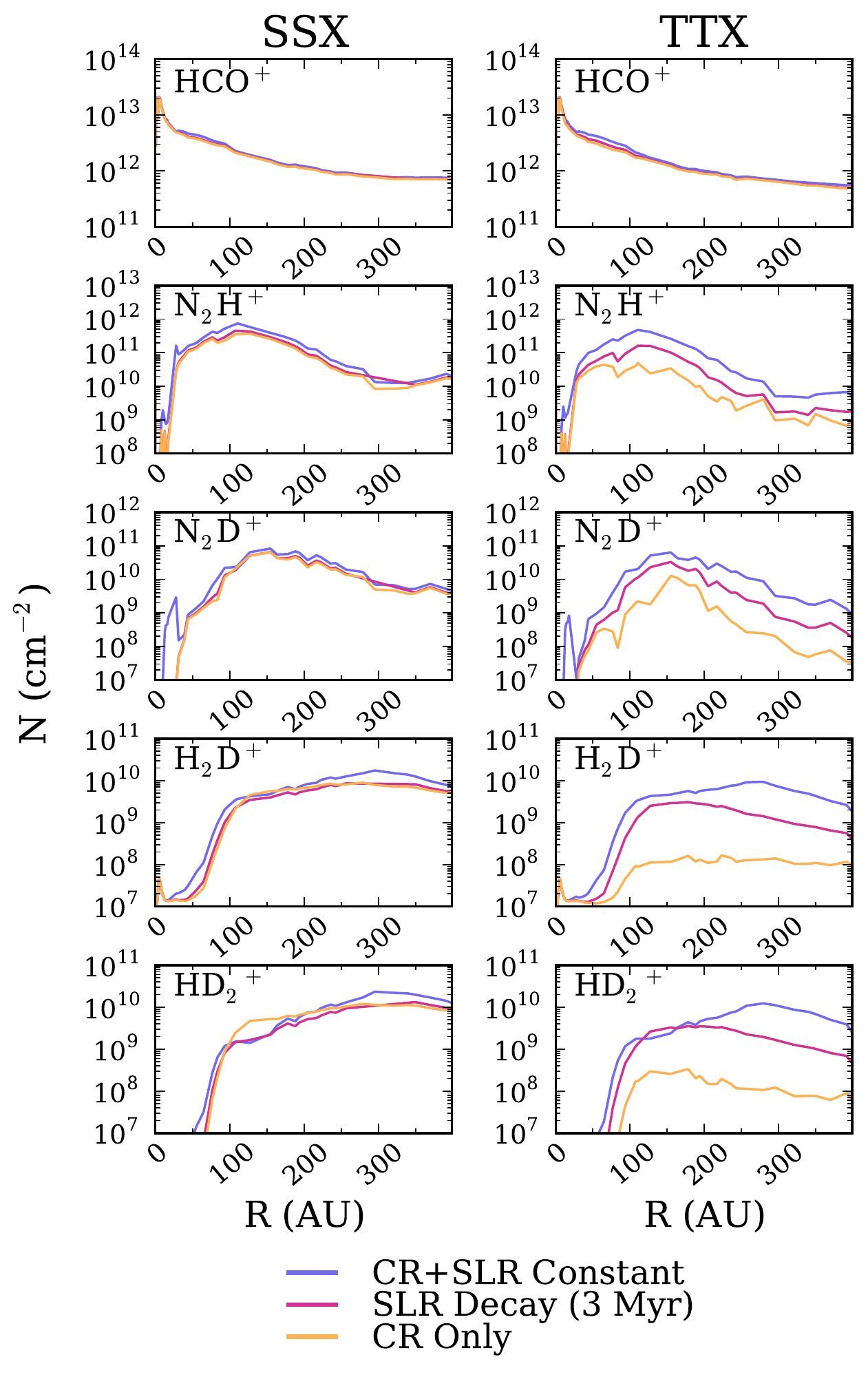}
\caption{Column density dependence on radionuclide ionization now including time dependence.  All curves are shown at $t=3$~Myr of chemical evolution.  The blue line shows a model with CRs and a fixed SLR ionization rate set by the initial value, no time decay included.  Magenta curves include a decay on the SLR ionization rate with a half-life of $t_{\rm{half}}\sim1.2$~Myr \citep{cleeves2013b}.  The yellow line shows the column density in the absence of SLR ionization.  The SSX model (as well as the higher CR ionization models, M02 and W98) do not depend on the time decay of SLRs as expected.  The TTX model is not buffered by CR ionization and is hence far more sensitive to the time decay of SLRs, except in the case of HCO$^+$.  \label{fig:radnuc}}
\end{figure}

Figure~\ref{fig:radnuc} shows the effect of time-decaying SLRs on molecular column densities for the two lowest CR ionization models, SSX and TTX, where the change is most significant.  The Figure~\ref{fig:radnuc} column densities are shown at $t=3$~Myr of time evolution instead of 1~Myr, which corresponds to a decrease of nearly an order of magnitude in the SLR ionization rate.  We show the model for fixed initial SLR abundances (blue), for CR only (orange), and a time-decaying SLR ionization rate (magenta). The inclusion of time-dependence for the SLRs {\em does not sensitively affect} the SSX model, which carries similar contribution from both CR and SLR ionization (at $t=0$~Myr). In other words, in the absence of SLRs, the CRs provide similar levels of ionization in the SSX case.  The CR ``absent'' TTX models (Figure~\ref{fig:radnuc}, right) are far more sensitive to the change in SLR rate, where the time decay changes the column densities of \ntdp, \htdp, and \hdtp\ by at least an order of magnitude.  Note the HCO$^+$ column is insensitive to differences in the SLR model assumptions, regardless of CR rate, as is expected by an X-ray dominated chemistry.

\subsection{Dependence on Disk Mass}\label{sec:mass}
The results of our study are for a particular disk model (\S\ref{sec:str}) and disk mass ($M_g=0.04$~M$_\odot$). Observed protoplanetary disks show significant diversity in mass, diameter, stellar properties and environment \citep{williamscieza}.  It is therefore interesting to quantify the sensitivity of these results to the specific choice of disk mass.  To test this scenario, we take our standard model and find the chemical signatures for a disk of half mass ($M_g=0.02$~M$_\odot$) and double mass ($M_g=0.08$~M$_\odot$).  To facilitate a simple comparison, we do not change the UV radiation field or temperature structure, but we do recompute the X-ray and CR ionization field.  In reality, an increase or decrease by a factor of $\sim2$ in density would change the disk opacity and would result in a correspondingly cooler (warmer) disk by $\lesssim10\%$ in dust temperature.  In the present section we focus on the abundances determined from the column densities; however, we note that the emission line ratios for a particular species will also reflect the change in local temperature due to the mass change.  Furthermore, a larger (smaller) mass also would make for a more (less) UV/X-ray shielded disk.  However, by changing a single parameter we may investigate, in this case, the role of more or less efficient ion-recombination and how this effect plays into the measured abundances (see \S\ref{sec:temperature} for temperature dependence).  We leave the geometrical parameters of the disk unchanged such that the disk density scales with the change in mass.  CRs are slightly more excluded by the higher gas column (not a large overall effect) and we do not consider SLRs here.  A higher disk mass will trap more SLR decay products prior to loss but it is only a small effect \citep[$\sim1.2\times$ more ionization;][]{cleeves2013b}.  

\begin{figure*}[ht!]
\centering
\includegraphics[width=0.93\textwidth]{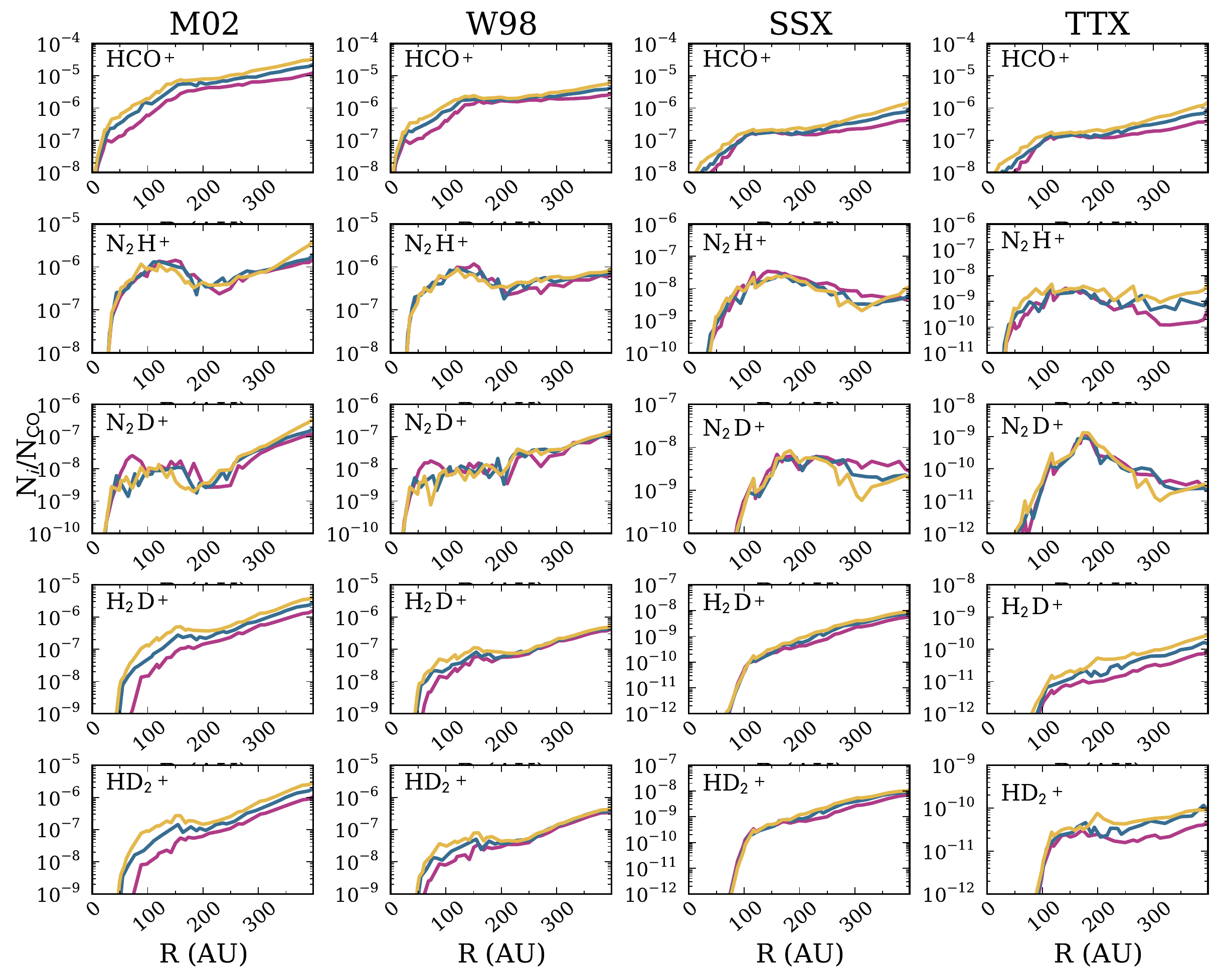}
\caption{Column-derived abundances of the indicated molecular ions, normalized to CO as a function of disk radius. Colors represent disk mass, where yellow, blue and magenta correspond to $0.5\times$, $1\times$, and $2\times$ disk gas mass models.  CR models are as indicated in the column headings.  Changes in the normalized column densities of a given molecular ion for different disk masses are, in general, far smaller than changes across different CR ionization rates.  Ionization decreases as one goes from left to right. See text for details and discussion. \label{fig:mass}}
\end{figure*}

In Figure~\ref{fig:mass}, we show the column of the indicated molecular ion, $N_i$, normalized to the column of model CO, $N_{\rm CO}$, which acts as our observable mass-reference.  Observations of optically thin lines provide their respective total (line-of-sight integrated) column densities. However, when one wants to determine relative abundances, typically to H$_2$, a normalization quantity is required.  The total gas mass can be inferred from millimeter dust emission or from molecular gas traced by CO, in both cases requiring a calibrated conversion factor.  Both methods have substantial caveats.  The millimeter-wave dust emission evolves as the underlying population of dust evolves via grain growth, making the dust-to-gas conversion factor a time-dependent quantity.  The CO abundance relative to H$_2$ likewise can be unreliable, both owing to freeze-out at low temperatures and chemical processing initiated by ionization \citep{bergin2014}, though this process may be slowed by vertical mixing \citep{furuya2014} and by grain growth through decreased surface area for freeze-out \citep{bergin2014}.  Ideally a less chemically reactive mass tracer like HD should be used for normalization, if available \citep{bergin2013}.  Nonetheless, given limited data on HD in protoplanetary disks, we provide column densities relative to the CO traced gas column, a far more widely available gas mass probe. In practice, CO column densities are extracted from optically thin isotopologues such as C$^{18}$O, however we emphasize that even C$^{18}$O has a minimum optical depth of $\tau\sim1$ at line center in this particular model (see Fig.~\ref{fig:tau}).  To determine the CO column observationally, one must consider either i) more rarified CO isotopes than C$^{18}$O or ii) emission restricted to the line-wings to ensure the line is thin.  In this analysis, the column densities of both the ions and of CO are taken directly from the chemical models.

In general, for the intermediate ionization rate models, W98 and SSX, the molecular ion abundance is not very sensitive to changes in the disk mass (formally density).  Most importantly, the abundances are far more sensitive to the CR ionization rate than they are to the disk mass.  Nonetheless, some variation does exist, where in Figure~\ref{fig:mass} the line thickness increases for increasing disk mass.  HCO$^+$ in particular shows a decrease in abundance with disk mass across all CR models.  These results can be understood by the same relation discussed in \S\ref{sec:xray},  where the ion abundance is proportional to $\sqrt{ \zeta / (\alpha n_{\rm{H_2}})}$.  In terms of column densities, the ratio of the ion-species to CO is then approximately given by $N_i/N_{\rm{CO}}\sim\chi_i N_{\rm{H_2}}/\chi_{\rm{CO}}N_{\rm{H_2}} = \chi_i/\chi_{\rm{CO}} = \chi_{\rm{CO}}^{-1}\sqrt{ \zeta / (\alpha n_{\rm{H_2}})}$.  For a more massive disk, $\zeta_{\rm XR}$ decreases due to increased gas opacity, and $n_{\rm H_2}$ increases as the mass increases.  Thus the quantity $\zeta / n_{\rm{H_2}}$ will decrease for increasing mass, reducing the ion abundance, explaining the general trend for all of the molecular ions considered here.   

Two special cases are the high M02 CR rate (leftmost column) and the low TTX CR rate (rightmost column).  Generally speaking, the M02 CR models tend to damp out the change in X-ray ionization, $\zeta_{\rm XR}$, leaving $\sqrt{n_{\rm H_2}^{-1}}$ as the dominant term.  M02 shows substantially more variation than can be attributed to changes in $n_{\rm H_2}$ or in X-ray flux, otherwise the variation would be seen in W98 as well.  This behavior can be understood by looking towards the denominator of the abundance ratio:  $\chi_{\rm{CO}}$.  As seen in the column density plots in Figure~\ref{fig:column}, the CO abundance is decreased due to chemical processing, and this deficit is reflected in the column integrated abundance, especially for  \htdp\ and \hdtp, which are far more spread out in abundance than the W98 and SSX models in Figure~\ref{fig:mass}.  

TTX is also a special case because the molecular ion abundances are very sensitive to the X-ray ionization rate, its only ionizing source.  Thus TTX is far more sensitive to the mass of the disk because $N_i/N_{\rm{CO}}$ depends sensitively on both $\zeta_{\rm XR}$ and $n_{\rm H_2}$.  Regardless, we note that the abundance spread over different mass models -- even for the TTX case -- is far less than the spread resulting from different ionization rates.  We thus conclude that ionization (X-ray, CR and SLRs) is the more important quantity regulating ion abundances, not disk gas mass.

\subsection{Dependence on Stellar Spectral Type}\label{sec:temperature}
\begin{figure*}[ht!]
\centering
\includegraphics[width=0.8\textwidth]{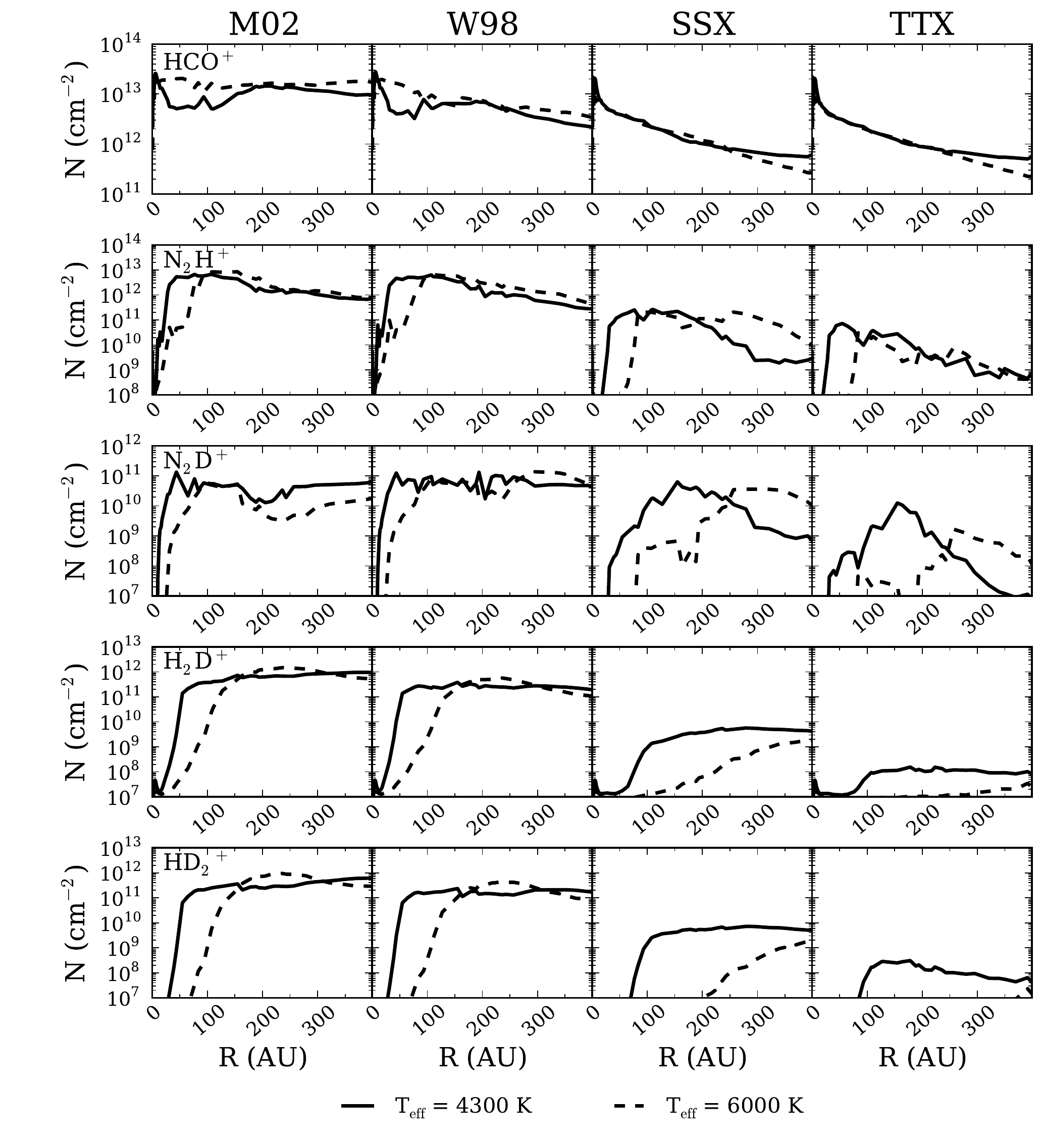}
\caption{Vertical column densities of select species as a function of stellar spectral type.  Solid lines correspond to the fiducial disk model with stellar effective temperature of $T_{\rm eff}=4300$~K.  Dashed lines show column densities for the warmer (more luminous) central star, $T_{\rm eff}=6000$~K.  See discussion in \S\ref{sec:temperature} for details. \label{fig:startemp}}
\end{figure*}

The chemical properties of disks are especially sensitive to the dust and gas temperatures, which are set by irradiation from the central star at radii beyond the inner $R\sim1$~AU, i.e., the region where accretion heating becomes significant.  The temperature in the warm molecular layer is especially important in regulating the observable gas-phase chemistry \citep{aikawa2002}.  To explore the chemical dependence on the temperature of the central star, we  recompute our baseline chemical model for a central star with T$_{\rm eff}=6000$~K, where all other structural parameters are held fixed.   At $R=15$~AU, the disk surface ($z/r=0.3$) temperature increases substantially from $T_d\sim100$~K to $\sim140$~K.  The warm molecular layer ($z/r=0.1$) increases from $T_d=40$~K to 55~K, and the midplane increases by $\Delta T_d=10$~K from 30~K to 40~K.  This substantial increase in temperature has a sizable effect on the abundance profiles as shown in Figure~\ref{fig:startemp}. 
Specifically, the increase in disk temperature pushes out the CO snow line, resulting in a larger abundance of CO in the inner disk, and a thicker layer of CO radially across the entire disk.  The HCO$^+$ column densities in Figure~\ref{fig:startemp} for the M02 and W98 (high CR flux) cases reflect the enhanced HCO$^+$ production from the additional CO.  However, the effect is mediated by the same chemical processing of CO as discussed in \S\ref{sec:chemres} and \S\ref{sec:mass}, where for the warmer star, CO processing occurs deeper in the dense layers of the disk, closer to the midplane, and thereby has a larger effect on the total column of CO than for the cooler star.  
HCO$^+$ for the low CR-ionization cases, SSX and TTX, shows a decrease in the outer disk, beyond $R>250$~AU, due to the combination of CO reprocessing happening deeper in the warmer disk and the simultaneous loss of non-thermal CR desorption, enabling rapid carbon sequestration from CO into other ices.

For all CR-ionization models, the column density of \nthp\ shows a decrease in the inner $R<100$~AU and a surplus beyond $R>150$~AU in the cold versus warm disk model.  The inner deficit is a direct result of the enhanced inner disk CO abundance and increased destruction rate. The outer enhancement follows from the increased gas-phase N$_2$ abundance in the warmer disk model where it would otherwise freeze out, thereby enabling N$_2$H$^+$ formation.  This process operates in tandem with the loss of CO in the dense gas due to reprocessing, reducing one of the primary destruction agents of N$_2$H$^+$.  The \ntdp\ column density shows similar morphological changes as \nthp\ where it is pushed further out radially; however, its abundance is simultaneously affected by the net reduction in deuterium-bearing species in the warm gas and the immediate loss of \htdp\ relative to H$_3^+$.  The \htdp\ column density has the same drop in the inner disk due to the increased CO abundance and overall warmer gas temperatures, though the change is less severe in the outer disk mainly because its precursor, H$_2$, does not freeze-out in either model.

The warmer disk model not only changes the chemical properties of the disk, but also the observational strategy necessary to experimentally determine the disk ionization processes, especially that of CRs.  In the cold disk case, all molecular ions considered here were able to distinguish between high (M02 and W98) and low (SSX and below) CR fluxes outside of $R>50$~AU; however, the column density of these species are less sensitive to changes at lower ionization rates due to freeze-out of CO and N$_2$.  For the warm disk, the CO freeze-out region is pushed further out radially and the HCO$^+$ column density becomes more sensitive to the CR ionization rate outside of the X-ray dominated region ($R>50$~AU).  At $R\sim175$~AU, the SSX and TTX CR ionization rates are indistinguishable for the cold disk model because of CO freeze-out, but in the warm disk case, there is a $\sim60\%$ difference between the SSX and TTX models.  While this difference is not sufficient to measure the ionization rate to great accuracy for this particular model, a warm disk around a more X-ray faint star (or a less X-ray permeable disk) may engender conditions favorable for optically thin HCO$^+$ isotopologue emission tracing {\em CR-dominated} layers down to low CR flux-levels.  \nthp\ and \ntdp\ remain sensitive observational tracers of the CR ionization rate at both high and low CR fluxes  beyond $R>100$~AU, where the warm disk acts to increase the ``dynamic range'' between the model column densities due to the reduced N$_2$ freeze-out. 
For example, the variation in \nthp\ column density between SSX and TTX models was approximately one order of magnitude or less in the fiducial cold disk (see Figure~\ref{fig:column}).  The warm disk increases the fractional difference to over two orders of magnitude in \nthp\ and the same behavior holds true for \ntdp.  Thus warmer disks may allow us to estimate more precisely the CR ionization rate by mediating the effects of freeze-out and increasing the overall column.  One potential caveat is that the radially larger ``snowline'' makes for a higher disk-averaged CO column and disk-integrated CO opacity compared to a cooler disk.  In this case, even rarer CO isotopes than C$^{18}$O may be necessary to interpret the HCO$^+$ column densities or to use as a proxy for gas mass.  It is also important to point out that the emission is not only sensitive to the column but also the emitting temperature, and therefore the particular transition or transitions targeted should also take into account the inherent temperature of the disk as estimated by the stellar luminosity.

\subsection{Disk Magnetic ``Opacity'' to Cosmic Ray Ionization}\label{sec:magnetic}
The CR contribution to disk ionization in both magnitude and scope is the least observationally constrained parameter in disks.  A cosmic ray ionization rate of $\zeta_{\rm{CR}}\gtrsim3\times10^{-17}$~s$^{-1}$, consistent with dense ISM values, was ruled out by \citet{chapillon2011} using observations of \htdp.  There are a few possible explanations for the low CR rates inferred.  The present work focuses on the possibility of wind-modulated {\em incident} CR ionization rates due to the presence of an analogue to the Solar System's heliosphere.  Within this paradigm we have ignored radial variations in the GCR rate that may include (i) a gradual increase (1\%/AU) in the CR rate with radius where modulation is weaker further out in the disk (Paper I) and (ii) an edge to the region of CR-modulation, i.e, a Heliopause.  As compared to the negative ionization gradient expected for SLR ionization, (i) would create a positive ionization gradient, though the two effects are similar ($\sim1$ order of magnitude) and may conspire to cancel each other out.  The other important radial effect is the extent of the wind-modulation zone, the ``T-Tauriosphere.'' We have assumed the disk is fully enclosed, but if winds can only punch out the inner tens of AU, perhaps by magnetic trapping of the winds \citep{turner2014}, the region of exclusion may be much smaller.  Alternatively, if the winds dominate, they could perhaps encircle a much larger region, hundreds to thousands of AU in size due to the high gas densities (and correspondingly higher ram pressures) of early stellar winds. 

Magnetic fields provide an additional source of ``opacity'' to CRs in two ways: (i) for funnel-shaped magnetic field configurations seen in protostars \citep[e.g.,][]{girart2006}, the magnetic field can mirror at most 50\% of the CRs away \citep[see also][]{padovani2011} and (ii) the presence of magnetic irregularities with size scales near the CR gyro-radius can scatter CRs \citep{cesarsky1978}.  Both effects can act in tandem, where magnetic irregularities on an hourglass magnetic field configuration can further enhance the fraction of mirrored particles \citep{fatuzzo2014}.   The role of CR exclusion by magnetic irregularities induced by disk gas turbulence has been explored in \citet{dolginov1994}.  It was found that modest magnetic field strengths with imposed turbulent irregularities significantly impeded the CR rate in the disk, such that only $20-30\%$ of CRs reach depths corresponding to the disk scale height at all radii beyond the inner few AU.  Including irregularities causes the CR ionization rate to decay very quickly with vertical depth towards the midplane \citep[see Eq.~(9) of][]{dolginov1994}, thus removing CRs entirely by the midplane.  Plugging in the numbers typical of our own disk model, we find that magnetic irregularities act to reduce the CR rate at the midplane by six orders of magnitude at 100~AU compared to the CR rate at the disk surface.   Furthermore, both winds {\em and} magnetic effects can operate simultaneously, such that the winds reduce the {\em incident} CR ionization rate and disk magnetic irregularities substantially curtail the CR propagation {\em internal } to the disk, much faster the classically assumed penetration depth of 100~g~cm$^-2$ \citep{umebayashi1981}.

The two scenarios, winds and ``magnetic opacity,'' act similarly to reduce the CR flux in the disk's midplane but where they differ is in the {\em surface}.  While we have touted the HCO$^+$ isotopologue emission as an excellent X-ray tracer, it is still sensitive to the higher CR ionization rate models considered here (with $\zeta_{\rm{CR}}\gtrsim10^{-17}$~s$^{-1}$, i.e., M02 and W98).  If winds are modulating the CR ionization, we would expect the HCO$^+$ emission to reflect a uniformly low CR rate regardless of height.  If magnetic irregularities dominate the attenuation of CRs,  then the CR ionization rate should be normal in the HCO$^+$ traced upper layers and absent in the midplane.  To conduct such an experiment however, requires reasonably good constraints on the stellar X-ray luminosity and distributions of density and temperature in the disk, as both of these are expected to affect the HCO$^+$ emission (\S\ref{sec:xray} and \S\ref{sec:mass}, respectively).  Alternatively, if neither of these effects are important, wind modulation nor magnetic effects, then the ionization should be $\zeta_{\rm{CR}}\gtrsim10^{-17}$~s$^{-1}$ in the outer disk and the more weakly emissive, difficult species to observe, like \htdp, can help provide constraints on midplane ionization.

\section{Discussion and Conclusions}
\label{sec:discussion}

Using a generic, observationally motivated model of a T Tauri protoplanetary disk, we present chemical abundance models and simulated submillimeter emission line observations that can be used as a blueprint to constrain the detailed ionization environment within the gas disk. In particular, sensitive ($\sim6h$) ALMA observations of nearby protoplanetary disks ($D\sim100$~pc) will readily be able to distinguish between systems with high, interstellar CR ionization levels ($\zeta_{\rm{CR}}\gtrsim10^{-17}$~s$^{-1}$) and those with low (sub-interstellar) ionization levels, though determining very low CR rates ($\zeta_{\rm{CR}}\lesssim10^{-20}$~s$^{-1}$) will be made difficult by weak emission from molecular tracers and X-rays providing a lower ionization limit that may hide the effects of CRs.
We emphasize that the chemical results presented here demonstrate {\em relative} trends across different ionization models and that the physical structure of the disk and properties of the star will determine how the chemical abundances measured directly map to ionization properties of the disk in an absolute sense.  Consequently, creating detailed models of particular sources with as many observational constraints as possible will be crucial to mapping out the ionization in detail for any particular system.

We highlight the molecular ions that are useful tracers of specific ionizing agents in dense gas,  e.g., the warm, X-ray irradiated molecular surface through HCO$^+$ or the cold, dense SLR and/or CR dominated midplane through \ntdp\ and \htdp.  Moreover, by isolating individual ionization sources, whether by central stellar processes or otherwise, we can use these results to observationally quantify the relative importance of each ionizing agent to the total ionization state of the disk gas.  Better constraints on the underlying ionization environment inform models of turbulence via the ionization-dependent magneto-rotational instability \citep{balbus1991} and models of disk chemistry via ion-neutral and grain surface reactions. In other words, dense gas ionization drives both the fundamental processes that govern both the formation (by potentially regulating turbulence-free ``dead zones'') and chemical make-up of planetary systems.

In particular, the least observationally constrained ionizing agents present in disks are those which dominate the midplane ionization -- CRs and SLRs.  More specifically, the CR rate incident on a protoplanetary disk is unknown and may be strongly modulated by winds and/or magnetic fields.  Without CRs, the total midplane ionization rate is reduced by at least one to two orders of magnitude depending upon radial location and the degree to which SLR ionization contributes, which becomes the primary midplane ionization source in the absence of CRs (see also Paper I).  High spatial resolution observations may furthermore reveal radial structure (e.g., gradients) in the CR rate with distance from the star, potentially belying the presence of an analogue heliosphere to that of the Solar System.  The molecular tracers outlined in this paper thus provide signposts of the presence or absence of important physical processes related to disk ionization.  In summary, the main results of this work are:

\begin{enumerate}
\item Chemical abundances of different molecular ions trace different ionizing agents (and regions) in our fiducial disk model.  The abundance variation is born out in their submillimeter emission, and thus the lines discussed here can be used as diagnostic observational tools of disk ionization.
\item Optically thin isotopes of HCO$^+$ trace primarily the incident X-ray ionizing flux, in particular X-rays with energies $E_{\rm XR}\sim5-7$~keV.  For a generally warmer disk, the CO snow line may exist further out, and in this scenario, radially extended (resolved) HCO$^+$ emission may be sensitive to midplane ionization sources, i.e., CRs and SLRs.
\item N$_2$H$^+$ is sensitive to both cold ionization processes (CRs and SLRs) and warm ($T>20$~K) gas ionization via stellar X-rays and is likely moderately optically thick inside $R\lesssim200$~AU ($\tau$ of a few).  This behavior makes for somewhat difficult direct interpretation, requiring detailed chemical/line emission models and multiple transitions.
\item We find that the cold gas ionization tracer \htdp\ becomes undetectable in a $6h$ full ALMA observation at $D=100$~pc for CR ionization rates below solar maximum (SSX models).  To detect the amount of \htdp\ present in our models at the level of $\zeta_{\rm{CR}}\sim10^{-19}$~s$^{-1}$, at least an order of magnitude more sensitive observations would be required, or a less distant disk such as TW Hya. 
\item \ntdp, rather than \htdp\ is a more sensitive tracer of midplane ionization due to CRs and SLRs and the least sensitive to stellar X-rays.  \ntdp\ can be used to measure ionization levels even at low CR rates ($\zeta_{\rm CR}\lesssim10^{-19}$~s$^{-1}$) and should be detectable with ALMA in a reasonable amount of time ($<10h$).  
\item Cosmic rays may be excluded by wind processes or internal/external magnetic processes (or both), but combined observations of surface ionization tracers such as HCO$^+$ isotopologues and midplane tracers like \ntdp\ will help illuminate the CR-exclusion mechanism that dominates (see \S\ref{sec:magnetic}).
\item Short-lived radionuclides are an important contributor to the ionization in disks, exceeding that of scattered stellar X-rays within the first few Myr.  If CRs are not present, \ntdp\ should still be detectable for young disks due to the SLR contribution, but its emission will fade over the disk lifetime. 
\item It will be very difficult to tell apart SLR ionization from CR ionization for measured midplane ionization rates of $\zeta\sim(1-10)\times10^{-19}$~s$^{-1}$. A negative ionization gradient with radius would indicate a SLR dominated chemistry (SLRs can escape the tenuous outer disk radii), while a flat or positive gradient (if the inner disk is very dense with $\Sigma_g>100$~g~cm$^{-2}$) may point to a CR dominated chemistry.
\item The mass-normalized column density of ions (with respect to HD, CO, or dust) is far more sensitive to the CR ionization rate than to the mass of the disk itself, where we have considered disk masses spanning the range $M_g=0.02-0.08$~M$_\odot$. 
\item Not all emission ``rings'' trace physical deficits in abundance or structure.
Large temperature gradients present in disks can result in low-J rotational lines peaking offset from the star due to de-excitation in warm gas, such as can happen with HCO$^+$ for high CR rates.
\end{enumerate}

This paper provides a viable starting point to study the ionization sources acting in circumstellar disks and their corresponding chemical signatures. However, this work must be carried forward in several ways. First, the ionization models used here are preliminary. We need to construct more detailed models for how CRs are suppressed by both T Tauri winds and magnetic field fluctuations; we also need improved assessments of the ionization rates provided by background cluster environments.  On another front, both existing and upcoming submillimeter facilities will provide important constraints on the actual ionization levels realized in these systems and will determine which molecules provide the most information. With improved theoretical and observational input, the chemical signatures considered here can then be revisited. In the end, we will thus obtain a good working understanding of both the relevant ionization processes and the chemical structure of planet forming disks. This information, in turn, can then be used to constrain disk evolution.  More specifically, the chemical structure of the disk determines the locations of both the dead zones (where MRI cannot operate) and chemical gradients in the gas and ice, and these structures greatly influence the accompanying processes of disk accretion and planet formation.

\noindent \acknowledgements{{\em Acknowledgements:} The authors thank the anonymous referee and editor for their comments and suggestions, which have greatly improved the manuscript.  This work was supported by NSF grant AST-1008800. }

\appendix
\section{Updated Short-Lived Radionuclide Rates}\label{app:rn}

We provide fits to the midplane ionization rate in \citet{cleeves2013b} as a function of vertical surface density and time.  We note, however, that the previous work assumed a somewhat short half-life for $^{60}$Fe, $t_{\rm half}=1.49\pm0.27$~Myr from \citet{kutschera1984}.  This value was revised in \citet{rugel2009} to be $t_{\rm half}=2.62\pm0.04$ for a larger $^{60}$Fe sample.  In the present work we recalculate the SLR ionization rate in the same method as \citet{cleeves2013b} adopting the revised $^{60}$Fe half-life.  The effect only becomes important after a disk life-time of 5~Myr.  At this evolutionary time, the scattered stellar X-rays begin to dominate over the SLR contribution.  Consequently, the net disk ionization properties should not be strongly sensitive to the specific $^{60}$Fe half-life used.  

\begin{figure}[ht!]
\centering
\includegraphics[width=0.55\textwidth]{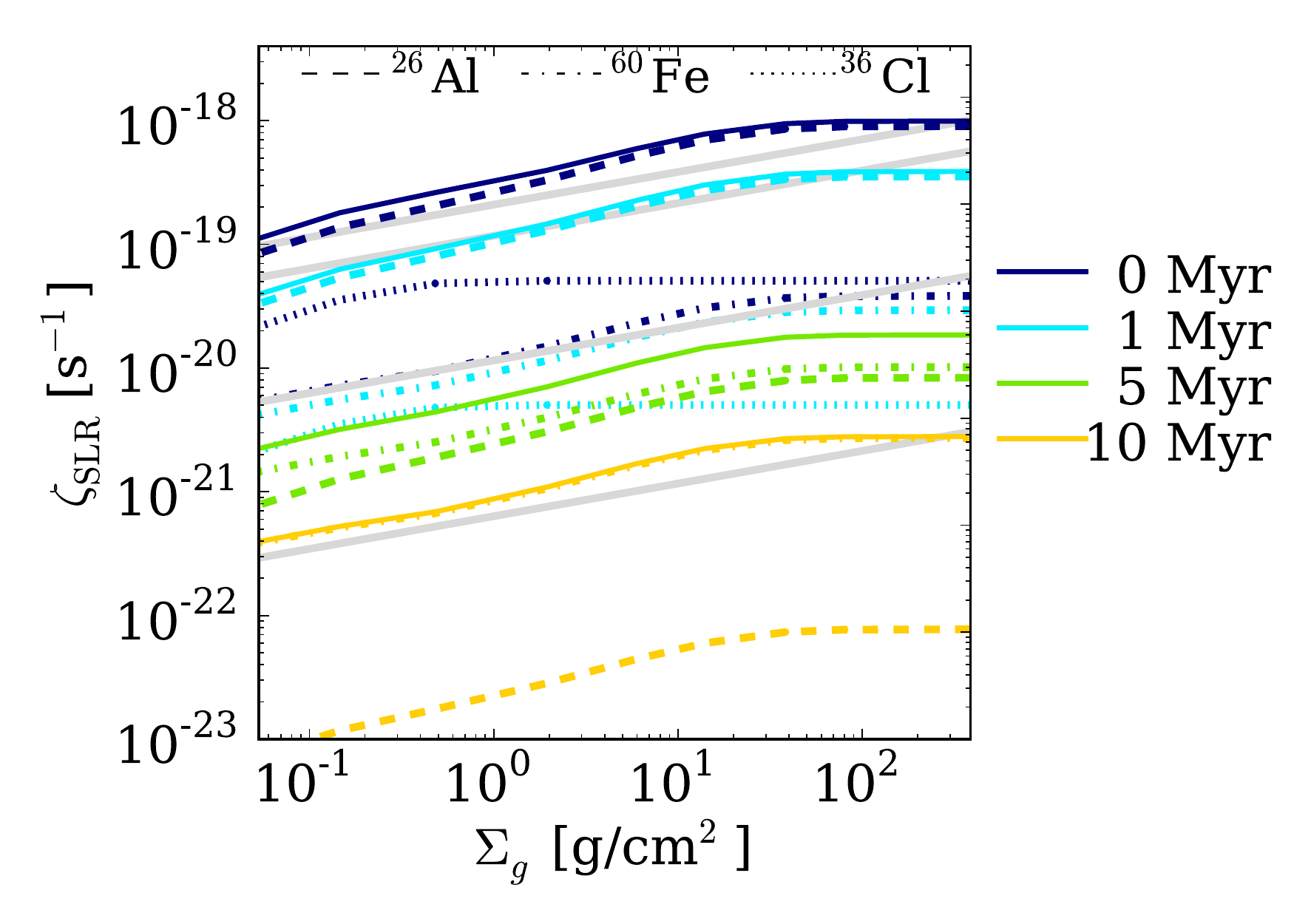}
\caption{Same as Fig.~4 of \citet{cleeves2013b}. Short-lived radionuclide H$_2$ ionization rate in the disk midplane as a function of {\em vertical} gas surface density and time (from the initial event suppling the SLR abundances, approximately the formation time of the disk).  Grey lines indicate the new fits to the values provided by Eq.~(\ref{eq:approxfit}).\label{newslr}}
\end{figure}

Nevertheless, in the interest of completeness, we provide updated SLR ionization rate calculations versus disk surface density in Figure~\ref{newslr}.   Colors indicate the time from ``disk formation,'' i.e., the epoch when the initial measured SLR abundances were set.  
We re-fit the ionization rate curves using a power law (grey lines), which are described by 
\be\label{eq:approxfit}\zeta_{\rm{H_2}}(r)=\left(2.1\times10^{-19}~{\rm{s}}^{-1}\right)
\left(1\over2\right)^{t/1.2}\left({\Sigma(r)\over{{\rm{g~cm^{-2}}}}}\right)^{0.26},\ee 
where time, $t$, is given in Myr.   This equation is an updated version of Equation~(30) in \citet{cleeves2013b}.  The change in the half-life of the SLR ensemble increases from $\sim1$~Myr to 1.2~Myr, meaning that SLRs can play an important role in the absence of CRs for a slightly longer span of the disk's lifetime.  We emphasize that the abundances of SLRs will vary from source to source.  This variation will change the leading coefficient of Eq.~(\ref{eq:approxfit} by at least of factor of two in both directions.


\begin{thebibliography}{129}
\expandafter\ifx\csname natexlab\endcsname\relax\def\natexlab#1{#1}\fi

\bibitem[{{Adams}(2010)}]{adams2010}
{Adams}, F.~C. 2010, \araa, 48, 47

\bibitem[{{Adams} {et~al.}(2014){Adams}, {Fatuzzo}, \& {Holden}}]{adams2014}
{Adams}, F.~C., {Fatuzzo}, M., \& {Holden}, L. 2014, \apj, 789, 86

\bibitem[{{Aikawa} \& {Herbst}(1999{\natexlab{a}})}]{aikawa1999}
{Aikawa}, Y. \& {Herbst}, E. 1999{\natexlab{a}}, \apj, 526, 314

\bibitem[{{Aikawa} \& {Herbst}(1999{\natexlab{b}})}]{aikawa1999a}
---. 1999{\natexlab{b}}, \aap, 351, 233

\bibitem[{{Aikawa} {et~al.}(2001){Aikawa}, {Ohashi}, {Inutsuka}, {Herbst}, \&
  {Takakuwa}}]{aikawa2001}
{Aikawa}, Y., {Ohashi}, N., {Inutsuka}, S.-i., {Herbst}, E., \& {Takakuwa}, S.
  2001, \apj, 552, 639

\bibitem[{{Aikawa} {et~al.}(2002){Aikawa}, {van Zadelhoff}, {van Dishoeck}, \&
  {Herbst}}]{aikawa2002}
{Aikawa}, Y., {van Zadelhoff}, G.~J., {van Dishoeck}, E.~F., \& {Herbst}, E.
  2002, \aap, 386, 622

\bibitem[{{Albertsson} {et~al.}(2014){Albertsson}, {Semenov}, \&
  {Henning}}]{albertsson2014}
{Albertsson}, T., {Semenov}, D., \& {Henning}, T. 2014, \apj, 784, 39

\bibitem[{{Amano} \& {Hirao}(2005)}]{amano2005}
{Amano}, T. \& {Hirao}, T. 2005, Journal of Molecular Spectroscopy, 233, 7

\bibitem[{{Anderson} {et~al.}(2013){Anderson}, {Bergin}, {Maret}, \&
  {Wakelam}}]{anderson2013}
{Anderson}, D.~E., {Bergin}, E.~A., {Maret}, S., \& {Wakelam}, V. 2013, \apj,
  779, 141

\bibitem[{{Anderson} {et~al.}(1977){Anderson}, {Dixon}, {Piltch}, {Saykally},
  {Szanto}, \& {Woods}}]{anderson1977}
{Anderson}, T.~G., {Dixon}, T.~A., {Piltch}, N.~D., {Saykally}, R.~J.,
  {Szanto}, P.~G., \& {Woods}, R.~C. 1977, \apjl, 216, L85

\bibitem[{{Andrews} {et~al.}(2011){Andrews}, {Wilner}, {Espaillat}, {Hughes},
  {Dullemond}, {McClure}, {Qi}, \& {Brown}}]{andrews2011}
{Andrews}, S.~M., {Wilner}, D.~J., {Espaillat}, C., {Hughes}, A.~M.,
  {Dullemond}, C.~P., {McClure}, M.~K., {Qi}, C., \& {Brown}, J.~M. 2011, \apj,
  732, 42

\bibitem[{{Asplund} {et~al.}(2009){Asplund}, {Grevesse}, {Sauval}, \&
  {Scott}}]{asplund2009}
{Asplund}, M., {Grevesse}, N., {Sauval}, A.~J., \& {Scott}, P. 2009, \araa, 47,
  481

\bibitem[{{Bachiller} {et~al.}(1990){Bachiller}, {Menten}, \& {del Rio
  Alvarez}}]{bachiller1990}
{Bachiller}, R., {Menten}, K.~M., \& {del Rio Alvarez}, S. 1990, \aap, 236, 461

\bibitem[{{Balbus} \& {Hawley}(1991)}]{balbus1991}
{Balbus}, S.~A. \& {Hawley}, J.~F. 1991, \apj, 376, 214

\bibitem[{{Bergin} {et~al.}(2002){Bergin}, {Alves}, {Huard}, \&
  {Lada}}]{bergin2002}
{Bergin}, E.~A., {Alves}, J., {Huard}, T., \& {Lada}, C.~J. 2002, \apjl, 570,
  L101

\bibitem[{{Bergin} {et~al.}(2014){Bergin}, {Cleeves}, {Crockett}, \&
  {Blake}}]{bergin2014}
{Bergin}, E.~A., {Cleeves}, L.~I., {Crockett}, N., \& {Blake}, G.~A. 2014,
  Faraday Discussions, Accepted

\bibitem[{{Bergin} {et~al.}(2013){Bergin}, {Cleeves}, {Gorti}, {Zhang},
  {Blake}, {Green}, {Andrews}, {Evans}, {Henning}, {{\"O}berg}, {Pontoppidan},
  {Qi}, {Salyk}, \& {van Dishoeck}}]{bergin2013}
{Bergin}, E.~A., {Cleeves}, L.~I., {Gorti}, U., {Zhang}, K., {Blake}, G.~A.,
  {Green}, J.~D., {Andrews}, S.~M., {Evans}, II, N.~J., {Henning}, T.,
  {{\"O}berg}, K., {Pontoppidan}, K., {Qi}, C., {Salyk}, C., \& {van Dishoeck},
  E.~F. 2013, \nat, 493, 644

\bibitem[{{Bergin} {et~al.}(1999){Bergin}, {Neufeld}, \&
  {Melnick}}]{bergin1999}
{Bergin}, E.~A., {Neufeld}, D.~A., \& {Melnick}, G.~J. 1999, \apjl, 510, L145

\bibitem[{{Bethell} \& {Bergin}(2011{\natexlab{a}})}]{bethell2011a}
{Bethell}, T.~J. \& {Bergin}, E.~A. 2011{\natexlab{a}}, \apj, 740, 7

\bibitem[{{Bethell} \& {Bergin}(2011{\natexlab{b}})}]{bethell2011b}
---. 2011{\natexlab{b}}, \apj, 739, 78

\bibitem[{{Brinch} \& {Hogerheijde}(2010)}]{brinch2010}
{Brinch}, C. \& {Hogerheijde}, M.~R. 2010, \aap, 523, A25

\bibitem[{{Calvet} {et~al.}(2005){Calvet}, {D'Alessio}, {Watson},
  {Franco-Hern{\'a}ndez}, {Furlan}, {Green}, {Sutter}, {Forrest}, {Hartmann},
  {Uchida}, {Keller}, {Sargent}, {Najita}, {Herter}, {Barry}, \&
  {Hall}}]{calvet2005}
{Calvet}, N., {D'Alessio}, P., {Watson}, D.~M., {Franco-Hern{\'a}ndez}, R.,
  {Furlan}, E., {Green}, J., {Sutter}, P.~M., {Forrest}, W.~J., {Hartmann}, L.,
  {Uchida}, K.~I., {Keller}, L.~D., {Sargent}, B., {Najita}, J., {Herter},
  T.~L., {Barry}, D.~J., \& {Hall}, P. 2005, \apjl, 630, L185

\bibitem[{{Ceccarelli} {et~al.}(2004){Ceccarelli}, {Dominik}, {Lefloch},
  {Caselli}, \& {Caux}}]{ceccarelli2004}
{Ceccarelli}, C., {Dominik}, C., {Lefloch}, B., {Caselli}, P., \& {Caux}, E.
  2004, \apjl, 607, L51

\bibitem[{{Cesarsky} \& {Volk}(1978)}]{cesarsky1978}
{Cesarsky}, C.~J. \& {Volk}, H.~J. 1978, \aap, 70, 367

\bibitem[{{Chapillon} {et~al.}(2011){Chapillon}, {Parise}, {Guilloteau}, \&
  {Du}}]{chapillon2011}
{Chapillon}, E., {Parise}, B., {Guilloteau}, S., \& {Du}, F. 2011, \aap, 533,
  A143

\bibitem[{{Cleeves} {et~al.}(2013{\natexlab{a}}){Cleeves}, {Adams}, \&
  {Bergin}}]{cleeves2013a}
{Cleeves}, L.~I., {Adams}, F.~C., \& {Bergin}, E.~A. 2013{\natexlab{a}}, \apj,
  772, 5

\bibitem[{{Cleeves} {et~al.}(2013{\natexlab{b}}){Cleeves}, {Adams}, {Bergin},
  \& {Visser}}]{cleeves2013b}
{Cleeves}, L.~I., {Adams}, F.~C., {Bergin}, E.~A., \& {Visser}, R.
  2013{\natexlab{b}}, \apj, 777, 28

\bibitem[{{Cohen} {et~al.}(2012){Cohen}, {Drake}, \& {K{\'o}ta}}]{cohen2012}
{Cohen}, O., {Drake}, J.~J., \& {K{\'o}ta}, J. 2012, \apj, 760, 85

\bibitem[{{Collings} {et~al.}(2004){Collings}, {Anderson}, {Chen}, {Dever},
  {Viti}, {Williams}, \& {McCoustra}}]{collings2004}
{Collings}, M.~P., {Anderson}, M.~A., {Chen}, R., {Dever}, J.~W., {Viti}, S.,
  {Williams}, D.~A., \& {McCoustra}, M.~R.~S. 2004, \mnras, 354, 1133

\bibitem[{{Consolmagno} \& {Jokipii}(1978)}]{consolmagno1978}
{Consolmagno}, G.~J. \& {Jokipii}, J.~R. 1978, The moon and the planets, 19,
  253

\bibitem[{{Dalgarno}(2006)}]{dalgarno2006}
{Dalgarno}, A. 2006, Proceedings of the National Academy of Science, 103, 12269

\bibitem[{{Dolginov} \& {Stepinski}(1994)}]{dolginov1994}
{Dolginov}, A.~Z. \& {Stepinski}, T.~F. 1994, \apj, 427, 377

\bibitem[{{Draine}(1978)}]{draine1978}
{Draine}, B.~T. 1978, \apjs, 36, 595

\bibitem[{{Dullemond} \& {Dominik}(2004)}]{dullemond2004}
{Dullemond}, C.~P. \& {Dominik}, C. 2004, \aap, 421, 1075

\bibitem[{{Dutrey} {et~al.}(1997){Dutrey}, {Guilloteau}, \&
  {Guelin}}]{dutrey1997}
{Dutrey}, A., {Guilloteau}, S., \& {Guelin}, M. 1997, \aap, 317, L55

\bibitem[{{Ercolano} \& {Glassgold}(2013)}]{ercolano2013}
{Ercolano}, B. \& {Glassgold}, A.~E. 2013, \mnras, 436, 3446

\bibitem[{{Fatuzzo} \& {Adams}(2008)}]{fatuzzo2008}
{Fatuzzo}, M. \& {Adams}, F.~C. 2008, \apj, 675, 1361

\bibitem[{{Fatuzzo} \& {Adams}(2014)}]{fatuzzo2014}
---. 2014, \apj, 787, 26

\bibitem[{{Favre} {et~al.}(2013){Favre}, {Cleeves}, {Bergin}, {Qi}, \&
  {Blake}}]{favre2013}
{Favre}, C., {Cleeves}, L.~I., {Bergin}, E.~A., {Qi}, C., \& {Blake}, G.~A.
  2013, \apjl, 776, L38

\bibitem[{{Feigelson} {et~al.}(1993){Feigelson}, {Casanova}, {Montmerle}, \&
  {Guibert}}]{feigelson1993}
{Feigelson}, E.~D., {Casanova}, S., {Montmerle}, T., \& {Guibert}, J. 1993,
  \apj, 416, 623

\bibitem[{{Feigelson} \& {Decampli}(1981)}]{feigelson1981}
{Feigelson}, E.~D. \& {Decampli}, W.~M. 1981, \apjl, 243, L89

\bibitem[{{Finocchi} \& {Gail}(1997)}]{finocchi1997}
{Finocchi}, F. \& {Gail}, H.-P. 1997, \aap, 327, 825

\bibitem[{{Flower}(1999)}]{flower1999}
{Flower}, D.~R. 1999, \mnras, 305, 651

\bibitem[{{Flower} {et~al.}(2006){Flower}, {Pineau Des For{\^e}ts}, \&
  {Walmsley}}]{flower2006}
{Flower}, D.~R., {Pineau Des For{\^e}ts}, G., \& {Walmsley}, C.~M. 2006, \aap,
  449, 621

\bibitem[{{Fogel} {et~al.}(2011){Fogel}, {Bethell}, {Bergin}, {Calvet}, \&
  {Semenov}}]{fogel2011}
{Fogel}, J.~K.~J., {Bethell}, T.~J., {Bergin}, E.~A., {Calvet}, N., \&
  {Semenov}, D. 2011, \apj, 726, 29

\bibitem[{{France} {et~al.}(2014){France}, {Schindhelm}, {Bergin}, {Roueff}, \&
  {Abgrall}}]{france2014}
{France}, K., {Schindhelm}, E., {Bergin}, E.~A., {Roueff}, E., \& {Abgrall}, H.
  2014, \apj, 784, 127

\bibitem[{{Fromang} {et~al.}(2002){Fromang}, {Terquem}, \&
  {Balbus}}]{fromang2002}
{Fromang}, S., {Terquem}, C., \& {Balbus}, S.~A. 2002, \mnras, 329, 18

\bibitem[{{Furuya} \& {Aikawa}(2014)}]{furuya2014}
{Furuya}, K. \& {Aikawa}, Y. 2014, ArXiv e-prints

\bibitem[{{Gammie}(1996)}]{gammie1996}
{Gammie}, C.~F. 1996, \apj, 457, 355

\bibitem[{{Garrod} \& {Pauly}(2011)}]{garrod2011}
{Garrod}, R.~T. \& {Pauly}, T. 2011, \apj, 735, 15

\bibitem[{{Geiss} \& {Gloeckler}(2003)}]{geiss2003}
{Geiss}, J. \& {Gloeckler}, G. 2003, \ssr, 106, 3

\bibitem[{{Girart} {et~al.}(2006){Girart}, {Rao}, \& {Marrone}}]{girart2006}
{Girart}, J.~M., {Rao}, R., \& {Marrone}, D.~P. 2006, Science, 313, 812

\bibitem[{{Glassgold}(1999)}]{glassgold1999}
{Glassgold}, A.~E. 1999, in IAU Symposium, Vol. 191, Asymptotic Giant Branch
  Stars, ed. T.~{Le Bertre}, A.~{Lebre}, \& C.~{Waelkens}, 337

\bibitem[{{Glassgold} {et~al.}(2004){Glassgold}, {Najita}, \&
  {Igea}}]{glassgold2004}
{Glassgold}, A.~E., {Najita}, J., \& {Igea}, J. 2004, \apj, 615, 972

\bibitem[{{Gu{\'e}lin}(1988)}]{guelin1988}
{Gu{\'e}lin}, M. 1988, in Topics in Molecular Organization and Engineering,
  Vol.~2, Molecules in Physics, Chemistry, and Biology, ed. J.~Maruani
  (Springer Netherlands), 175--187

\bibitem[{{Guilloteau} {et~al.}(2006){Guilloteau}, {Pi{\'e}tu}, {Dutrey}, \&
  {Gu{\'e}lin}}]{guilloteau2006}
{Guilloteau}, S., {Pi{\'e}tu}, V., {Dutrey}, A., \& {Gu{\'e}lin}, M. 2006,
  \aap, 448, L5

\bibitem[{{Habing}(1968)}]{habing1968}
{Habing}, H.~J. 1968, \bain, 19, 421

\bibitem[{{Harries}(2000)}]{harries2000}
{Harries}, T.~J. 2000, \mnras, 315, 722

\bibitem[{{Harries} {et~al.}(2004){Harries}, {Monnier}, {Symington}, \&
  {Kurosawa}}]{harries2004}
{Harries}, T.~J., {Monnier}, J.~D., {Symington}, N.~H., \& {Kurosawa}, R. 2004,
  \mnras, 350, 565

\bibitem[{{Hasegawa} {et~al.}(1992){Hasegawa}, {Herbst}, \& {Leung}}]{hhl}
{Hasegawa}, T.~I., {Herbst}, E., \& {Leung}, C.~M. 1992, \apjs, 82, 167

\bibitem[{{Hatchell} {et~al.}(1998){Hatchell}, {Thompson}, {Millar}, \&
  {MacDonald}}]{hatchell1998}
{Hatchell}, J., {Thompson}, M.~A., {Millar}, T.~J., \& {MacDonald}, G.~H. 1998,
  \aap, 338, 713

\bibitem[{{Herbst} \& {Klemperer}(1973)}]{herbst1973}
{Herbst}, E. \& {Klemperer}, W. 1973, \apj, 185, 505

\bibitem[{{Herczeg} {et~al.}(2002){Herczeg}, {Linsky}, {Valenti},
  {Johns-Krull}, \& {Wood}}]{herczeg2002}
{Herczeg}, G.~J., {Linsky}, J.~L., {Valenti}, J.~A., {Johns-Krull}, C.~M., \&
  {Wood}, B.~E. 2002, \apj, 572, 310

\bibitem[{{Herczeg} {et~al.}(2004){Herczeg}, {Wood}, {Linsky}, {Valenti}, \&
  {Johns-Krull}}]{herczeg2004}
{Herczeg}, G.~J., {Wood}, B.~E., {Linsky}, J.~L., {Valenti}, J.~A., \&
  {Johns-Krull}, C.~M. 2004, \apj, 607, 369

\bibitem[{{Hughes} {et~al.}(2011){Hughes}, {Wilner}, {Andrews}, {Qi}, \&
  {Hogerheijde}}]{hughes2011}
{Hughes}, A.~M., {Wilner}, D.~J., {Andrews}, S.~M., {Qi}, C., \& {Hogerheijde},
  M.~R. 2011, \apj, 727, 85

\bibitem[{{Hughes} {et~al.}(2008){Hughes}, {Wilner}, {Qi}, \&
  {Hogerheijde}}]{hughes2008}
{Hughes}, A.~M., {Wilner}, D.~J., {Qi}, C., \& {Hogerheijde}, M.~R. 2008, \apj,
  678, 1119

\bibitem[{{Hugo} {et~al.}(2009){Hugo}, {Asvany}, \& {Schlemmer}}]{hugo2009}
{Hugo}, E., {Asvany}, O., \& {Schlemmer}, S. 2009, \jcp, 130, 164302

\bibitem[{{Igea} \& {Glassgold}(1999)}]{igea1999}
{Igea}, J. \& {Glassgold}, A.~E. 1999, \apj, 518, 848

\bibitem[{{Indriolo} \& {McCall}(2012)}]{indriolo2012}
{Indriolo}, N. \& {McCall}, B.~J. 2012, \apj, 745, 91

\bibitem[{{Jonkheid} {et~al.}(2007){Jonkheid}, {Dullemond}, {Hogerheijde}, \&
  {van Dishoeck}}]{jonkheid2007}
{Jonkheid}, B., {Dullemond}, C.~P., {Hogerheijde}, M.~R., \& {van Dishoeck},
  E.~F. 2007, \aap, 463, 203

\bibitem[{{J{\o}rgensen} {et~al.}(2004){J{\o}rgensen}, {Sch{\"o}ier}, \& {van
  Dishoeck}}]{jorgensen2004}
{J{\o}rgensen}, J.~K., {Sch{\"o}ier}, F.~L., \& {van Dishoeck}, E.~F. 2004,
  \aap, 416, 603

\bibitem[{{Joseph} {et~al.}(1986){Joseph}, {Snow}, {Seab}, \&
  {Crutcher}}]{joseph1986}
{Joseph}, C.~L., {Snow}, Jr., T.~P., {Seab}, C.~G., \& {Crutcher}, R.~M. 1986,
  \apj, 309, 771

\bibitem[{{Jura} {et~al.}(2013){Jura}, {Xu}, \& {Young}}]{jura2013}
{Jura}, M., {Xu}, S., \& {Young}, E.~D. 2013, \apjl, 775, L41

\bibitem[{{Kahane} {et~al.}(1992){Kahane}, {Cernicharo}, {Gomez-Gonzalez}, \&
  {Guelin}}]{kahane1992}
{Kahane}, C., {Cernicharo}, J., {Gomez-Gonzalez}, J., \& {Guelin}, M. 1992,
  \aap, 256, 235

\bibitem[{{Keene} {et~al.}(1998){Keene}, {Schilke}, {Kooi}, {Lis}, {Mehringer},
  \& {Phillips}}]{keene1998}
{Keene}, J., {Schilke}, P., {Kooi}, J., {Lis}, D.~C., {Mehringer}, D.~M., \&
  {Phillips}, T.~G. 1998, \apjl, 494, L107

\bibitem[{{Kurosawa} {et~al.}(2004){Kurosawa}, {Harries}, {Bate}, \&
  {Symington}}]{kurosawa2004}
{Kurosawa}, R., {Harries}, T.~J., {Bate}, M.~R., \& {Symington}, N.~H. 2004,
  \mnras, 351, 1134

\bibitem[{{Kutschera} {et~al.}(1984){Kutschera}, {Billquist}, {Frekers},
  {Henning}, {Jensen}, {Xiuzeng}, {Pardo}, {Paul}, {Rehm}, {Smither}, {Yntema},
  \& {Mausner}}]{kutschera1984}
{Kutschera}, W., {Billquist}, P.~J., {Frekers}, D., {Henning}, W., {Jensen},
  K.~J., {Xiuzeng}, M., {Pardo}, R., {Paul}, M., {Rehm}, K.~E., {Smither},
  R.~K., {Yntema}, J.~L., \& {Mausner}, L.~F. 1984, Nuclear Instruments and
  Methods in Physics Research B, 5, 430

\bibitem[{{Langer} {et~al.}(1996){Langer}, {Castets}, \&
  {Lefloch}}]{langer1996}
{Langer}, W.~D., {Castets}, A., \& {Lefloch}, B. 1996, \apjl, 471, L111

\bibitem[{{Lee} \& {Bergin}(2014)}]{lee2014}
{Lee}, J.-E. \& {Bergin}, E.~A. 2014, ApJ, submitted.

\bibitem[{{MacPherson} {et~al.}(1995){MacPherson}, {Davis}, \&
  {Zinner}}]{macpherson1995}
{MacPherson}, G.~J., {Davis}, A.~M., \& {Zinner}, E.~K. 1995, Meteoritics, 30,
  365

\bibitem[{{Maret} \& {Bergin}(2007)}]{maret2007}
{Maret}, S. \& {Bergin}, E.~A. 2007, \apj, 664, 956

\bibitem[{{Matsumura} \& {Pudritz}(2003)}]{matsumura2003}
{Matsumura}, S. \& {Pudritz}, R.~E. 2003, \apj, 598, 645

\bibitem[{{Millar} {et~al.}(1989){Millar}, {Bennett}, \& {Herbst}}]{millar1989}
{Millar}, T.~J., {Bennett}, A., \& {Herbst}, E. 1989, \apj, 340, 906

\bibitem[{{Moskalenko} {et~al.}(2002){Moskalenko}, {Strong}, {Ormes}, \&
  {Potgieter}}]{moskalenko2002}
{Moskalenko}, I.~V., {Strong}, A.~W., {Ormes}, J.~F., \& {Potgieter}, M.~S.
  2002, \apj, 565, 280

\bibitem[{{M{\"u}ller} {et~al.}(2005){M{\"u}ller}, {Schl{\"o}der}, {Stutzki},
  \& {Winnewisser}}]{muller2005}
{M{\"u}ller}, H.~S.~P., {Schl{\"o}der}, F., {Stutzki}, S., \& {Winnewisser}, G.
  2005, Journal of Molecular Structure, 742, 215

\bibitem[{{M{\"u}ller} {et~al.}(2001){M{\"u}ller}, {Thorwirth}, {Roth}, \&
  {Winnewisser}}]{muller2001}
{M{\"u}ller}, H.~S.~P., {Thorwirth}, S., {Roth}, D.~A., \& {Winnewisser}, G.
  2001, \aap, 370, L49

\bibitem[{{{\"O}berg} {et~al.}(2010){{\"O}berg}, {Qi}, {Fogel}, {Bergin},
  {Andrews}, {Espaillat}, {van Kempen}, {Wilner}, \& {Pascucci}}]{oberg2010}
{{\"O}berg}, K.~I., {Qi}, C., {Fogel}, J.~K.~J., {Bergin}, E.~A., {Andrews},
  S.~M., {Espaillat}, C., {van Kempen}, T.~A., {Wilner}, D.~J., \& {Pascucci},
  I. 2010, \apj, 720, 480

\bibitem[{{{\"O}berg} {et~al.}(2011{\natexlab{a}}){{\"O}berg}, {Qi}, {Fogel},
  {Bergin}, {Andrews}, {Espaillat}, {Wilner}, {Pascucci}, \&
  {Kastner}}]{oberg2011b}
{{\"O}berg}, K.~I., {Qi}, C., {Fogel}, J.~K.~J., {Bergin}, E.~A., {Andrews},
  S.~M., {Espaillat}, C., {Wilner}, D.~J., {Pascucci}, I., \& {Kastner}, J.~H.
  2011{\natexlab{a}}, \apj, 734, 98

\bibitem[{{{\"O}berg} {et~al.}(2011{\natexlab{b}}){{\"O}berg}, {Qi}, {Wilner},
  \& {Andrews}}]{oberg2011a}
{{\"O}berg}, K.~I., {Qi}, C., {Wilner}, D.~J., \& {Andrews}, S.~M.
  2011{\natexlab{b}}, \apj, 743, 152

\bibitem[{{{\"O}berg} {et~al.}(2005){{\"O}berg}, {van Broekhuizen}, {Fraser},
  {Bisschop}, {van Dishoeck}, \& {Schlemmer}}]{oberg2005}
{{\"O}berg}, K.~I., {van Broekhuizen}, F., {Fraser}, H.~J., {Bisschop}, S.~E.,
  {van Dishoeck}, E.~F., \& {Schlemmer}, S. 2005, \apjl, 621, L33

\bibitem[{{Padovani} \& {Galli}(2011)}]{padovani2011}
{Padovani}, M. \& {Galli}, D. 2011, \aap, 530, A109

\bibitem[{{Pagani} {et~al.}(2009){Pagani}, {Vastel}, {Hugo}, {Kokoouline},
  {Greene}, {Bacmann}, {Bayet}, {Ceccarelli}, {Peng}, \&
  {Schlemmer}}]{pagani2009}
{Pagani}, L., {Vastel}, C., {Hugo}, E., {Kokoouline}, V., {Greene}, C.~H.,
  {Bacmann}, A., {Bayet}, E., {Ceccarelli}, C., {Peng}, R., \& {Schlemmer}, S.
  2009, \aap, 494, 623

\bibitem[{{Perryman} {et~al.}(1997){Perryman}, {Lindegren}, {Kovalevsky},
  {Hoeg}, {Bastian}, {Bernacca}, {Cr{\'e}z{\'e}}, {Donati}, {Grenon},
  {Grewing}, {van Leeuwen}, {van der Marel}, {Mignard}, {Murray}, {Le Poole},
  {Schrijver}, {Turon}, {Arenou}, {Froeschl{\'e}}, \&
  {Petersen}}]{perryman1997}
{Perryman}, M.~A.~C., {Lindegren}, L., {Kovalevsky}, J., {Hoeg}, E., {Bastian},
  U., {Bernacca}, P.~L., {Cr{\'e}z{\'e}}, M., {Donati}, F., {Grenon}, M.,
  {Grewing}, M., {van Leeuwen}, F., {van der Marel}, H., {Mignard}, F.,
  {Murray}, C.~A., {Le Poole}, R.~S., {Schrijver}, H., {Turon}, C., {Arenou},
  F., {Froeschl{\'e}}, M., \& {Petersen}, C.~S. 1997, \aap, 323, L49

\bibitem[{{Pickett} {et~al.}(1998){Pickett}, {Poynter}, {Cohen}, {Delitsky},
  {Pearson}, \& {M{\"u}ller}}]{pickett1998}
{Pickett}, H.~M., {Poynter}, R.~L., {Cohen}, E.~A., {Delitsky}, M.~L.,
  {Pearson}, J.~C., \& {M{\"u}ller}, H.~S.~P. 1998, \jqsrt, 60, 883

\bibitem[{{Pinte} {et~al.}(2009){Pinte}, {Harries}, {Min}, {Watson},
  {Dullemond}, {Woitke}, {M{\'e}nard}, \& {Dur{\'a}n-Rojas}}]{pinte2009}
{Pinte}, C., {Harries}, T.~J., {Min}, M., {Watson}, A.~M., {Dullemond}, C.~P.,
  {Woitke}, P., {M{\'e}nard}, F., \& {Dur{\'a}n-Rojas}, M.~C. 2009, \aap, 498,
  967

\bibitem[{{Prantzos} {et~al.}(1996){Prantzos}, {Aubert}, \&
  {Audouze}}]{prantzos1996}
{Prantzos}, N., {Aubert}, O., \& {Audouze}, J. 1996, \aap, 309, 760

\bibitem[{{Qi} {et~al.}(2008){Qi}, {Wilner}, {Aikawa}, {Blake}, \&
  {Hogerheijde}}]{qi2008}
{Qi}, C., {Wilner}, D.~J., {Aikawa}, Y., {Blake}, G.~A., \& {Hogerheijde},
  M.~R. 2008, \apj, 681, 1396

\bibitem[{{Roberts} {et~al.}(2004){Roberts}, {Herbst}, \&
  {Millar}}]{roberts2004}
{Roberts}, H., {Herbst}, E., \& {Millar}, T.~J. 2004, \aap, 424, 905

\bibitem[{{Roberts} \& {Millar}(2000)}]{roberts2000}
{Roberts}, H. \& {Millar}, T.~J. 2000, \aap, 361, 388

\bibitem[{{Rugel} {et~al.}(2009){Rugel}, {Faestermann}, {Knie}, {Korschinek},
  {Poutivtsev}, {Schumann}, {Kivel}, {G{\"u}nther-Leopold}, {Weinreich}, \&
  {Wohlmuther}}]{rugel2009}
{Rugel}, G., {Faestermann}, T., {Knie}, K., {Korschinek}, G., {Poutivtsev}, M.,
  {Schumann}, D., {Kivel}, N., {G{\"u}nther-Leopold}, I., {Weinreich}, R., \&
  {Wohlmuther}, M. 2009, Physical Review Letters, 103, 072502

\bibitem[{{Rydbeck} {et~al.}(1980){Rydbeck}, {Hjalmarson}, {Rydbeck}, {Ellder},
  {Kollberg}, \& {Irvine}}]{rydbeck1980}
{Rydbeck}, O.~E.~H., {Hjalmarson}, A., {Rydbeck}, G., {Ellder}, J., {Kollberg},
  E., \& {Irvine}, W.~M. 1980, \apjl, 235, L171

\bibitem[{{Saito} {et~al.}(1985){Saito}, {Kawaguchi}, \& {Hirota}}]{saito1985}
{Saito}, S., {Kawaguchi}, K., \& {Hirota}, E. 1985, \jcp, 82, 45

\bibitem[{{Sastry} {et~al.}(1981){Sastry}, {Helminger}, {Herbst}, \& {De
  Lucia}}]{sastry1981}
{Sastry}, K.~V.~L.~N., {Helminger}, P., {Herbst}, E., \& {De Lucia}, F.~C.
  1981, Chemical Physics Letters, 84, 286

\bibitem[{{Schindhelm} {et~al.}(2012){Schindhelm}, {France}, {Herczeg},
  {Bergin}, {Yang}, {Brown}, {Brown}, {Linsky}, \& {Valenti}}]{schindhelm2012}
{Schindhelm}, E., {France}, K., {Herczeg}, G.~J., {Bergin}, E., {Yang}, H.,
  {Brown}, A., {Brown}, J.~M., {Linsky}, J.~L., \& {Valenti}, J. 2012, \apjl,
  756, L23

\bibitem[{{Sch{\"o}ier} {et~al.}(2005){Sch{\"o}ier}, {van der Tak}, {van
  Dishoeck}, \& {Black}}]{schoier2005}
{Sch{\"o}ier}, F.~L., {van der Tak}, F.~F.~S., {van Dishoeck}, E.~F., \&
  {Black}, J.~H. 2005, \aap, 432, 369

\bibitem[{{Schramm}(1971)}]{schramm1971}
{Schramm}, D.~N. 1971, \apss, 13, 249

\bibitem[{{Semenov} \& {Wiebe}(2011)}]{semenov2011}
{Semenov}, D. \& {Wiebe}, D. 2011, \apjs, 196, 25

\bibitem[{{Semenov} {et~al.}(2004){Semenov}, {Wiebe}, \&
  {Henning}}]{semenov2004}
{Semenov}, D., {Wiebe}, D., \& {Henning}, T. 2004, \aap, 417, 93

\bibitem[{{Smith} {et~al.}(2004){Smith}, {Herbst}, \& {Chang}}]{smith2004}
{Smith}, I.~W.~M., {Herbst}, E., \& {Chang}, Q. 2004, \mnras, 350, 323

\bibitem[{{Svensmark}(2006)}]{svensmark2006}
{Svensmark}, H. 2006, Astronomische Nachrichten, 327, 871

\bibitem[{Tange(2011)}]{Tange2011a}
Tange, O. 2011, ;login: The USENIX Magazine, 36, 42

\bibitem[{{Telleschi} {et~al.}(2007){Telleschi}, {G{\"u}del}, {Briggs},
  {Audard}, \& {Scelsi}}]{telleschi2007}
{Telleschi}, A., {G{\"u}del}, M., {Briggs}, K.~R., {Audard}, M., \& {Scelsi},
  L. 2007, \aap, 468, 443

\bibitem[{{Thi} {et~al.}(2004){Thi}, {van Zadelhoff}, \& {van
  Dishoeck}}]{thi2004}
{Thi}, W.-F., {van Zadelhoff}, G.-J., \& {van Dishoeck}, E.~F. 2004, \aap, 425,
  955

\bibitem[{{Tieftrunk} {et~al.}(1994){Tieftrunk}, {Pineau des Forets},
  {Schilke}, \& {Walmsley}}]{tieftrunk1994}
{Tieftrunk}, A., {Pineau des Forets}, G., {Schilke}, P., \& {Walmsley}, C.~M.
  1994, \aap, 289, 579

\bibitem[{{Tielens} \& {Hagen}(1982)}]{tielens1982}
{Tielens}, A.~G.~G.~M. \& {Hagen}, W. 1982, \aap, 114, 245

\bibitem[{{Turner} {et~al.}(2014){Turner}, {Fromang}, {Gammie}, {Klahr},
  {Lesur}, {Wardle}, \& {Bai}}]{turner2014}
{Turner}, N.~J., {Fromang}, S., {Gammie}, C., {Klahr}, H., {Lesur}, G.,
  {Wardle}, M., \& {Bai}, X.-N. 2014, ArXiv e-prints

\bibitem[{{Umebayashi} {et~al.}(2013){Umebayashi}, {Katsuma}, \&
  {Nomura}}]{umebayashi2013}
{Umebayashi}, T., {Katsuma}, N., \& {Nomura}, H. 2013, \apj, 764, 104

\bibitem[{{Umebayashi} \& {Nakano}(1981)}]{umebayashi1981}
{Umebayashi}, T. \& {Nakano}, T. 1981, \pasj, 33, 617

\bibitem[{{Umebayashi} \& {Nakano}(2009)}]{umebayashi2009}
---. 2009, \apj, 690, 69

\bibitem[{{van Dishoeck} {et~al.}(2003){van Dishoeck}, {Thi}, \& {van
  Zadelhoff}}]{vandishoeck2003}
{van Dishoeck}, E.~F., {Thi}, W.-F., \& {van Zadelhoff}, G.-J. 2003, \aap, 400,
  L1

\bibitem[{{Vasileiadis} {et~al.}(2013){Vasileiadis}, {Nordlund}, \&
  {Bizzarro}}]{vasileiadis2013}
{Vasileiadis}, A., {Nordlund}, {\AA}., \& {Bizzarro}, M. 2013, \apjl, 769, L8

\bibitem[{{Wakelam} {et~al.}(2004){Wakelam}, {Caselli}, {Ceccarelli}, {Herbst},
  \& {Castets}}]{wakelam2004}
{Wakelam}, V., {Caselli}, P., {Ceccarelli}, C., {Herbst}, E., \& {Castets}, A.
  2004, \aap, 422, 159

\bibitem[{{Wakelam} {et~al.}(2011){Wakelam}, {Hersant}, \&
  {Herpin}}]{wakelam2011}
{Wakelam}, V., {Hersant}, F., \& {Herpin}, F. 2011, \aap, 529, A112

\bibitem[{{Walmsley} {et~al.}(2004){Walmsley}, {Flower}, \& {Pineau des
  For{\^e}ts}}]{walmsley2004}
{Walmsley}, C.~M., {Flower}, D.~R., \& {Pineau des For{\^e}ts}, G. 2004, \aap,
  418, 1035

\bibitem[{{Wasserburg} {et~al.}(2006){Wasserburg}, {Busso}, {Gallino}, \&
  {Nollett}}]{wasserburg2006}
{Wasserburg}, G.~J., {Busso}, M., {Gallino}, R., \& {Nollett}, K.~M. 2006,
  Nuclear Physics A, 777, 5

\bibitem[{{Webber}(1998)}]{webber1998}
{Webber}, W.~R. 1998, \apj, 506, 329

\bibitem[{{Williams} \& {Cieza}(2011)}]{williamscieza}
{Williams}, J.~P. \& {Cieza}, L.~A. 2011, \araa, 49, 67

\bibitem[{{Wolcott-Green} \& {Haiman}(2011)}]{wolcottgreen2011}
{Wolcott-Green}, J. \& {Haiman}, Z. 2011, \mnras, 412, 2603

\bibitem[{{Yonezu} {et~al.}(2009){Yonezu}, {Matsushima}, {Moriwaki}, {Takagi},
  \& {Amano}}]{yonezu2009}
{Yonezu}, T., {Matsushima}, F., {Moriwaki}, Y., {Takagi}, K., \& {Amano}, T.
  2009, Journal of Molecular Spectroscopy, 256, 238

\end{thebibliography}
\end{document}